\documentclass[aps,preprint,nofootinbib,preprintnumbers,superscriptaddress]{revtex4-1}
\usepackage{graphicx}
\usepackage{dcolumn}
\usepackage{bm}
\usepackage{amsmath}
\usepackage{amsfonts} 
\usepackage{latexsym}
\usepackage{bbm}
\usepackage{color}
\usepackage{amssymb}
\usepackage{amsthm}
\usepackage{epsf}
\usepackage{epsfig}
\usepackage{caption3}
\usepackage{appendix}
\usepackage{mathtools}
\usepackage{cases}
\usepackage{subfigure}
\usepackage{multirow}
\makeatletter

\newcommand{\Rmnum}[1]{\expandafter\@slowromancap\romannumeral #1@}
\makeatother

\makeatletter
\newsavebox\myboxA
\newsavebox\myboxB
\newlength\mylenA
\newcommand*\xoverline[2][0.75]{%
	\sbox{\myboxA}{$\m@th#2$}%
	\setbox\myboxB\null
	\ht\myboxB=\ht\myboxA%
	\dp\myboxB=\dp\myboxA%
	\wd\myboxB=#1\wd\myboxA
	\sbox\myboxB{$\m@th\overline{\copy\myboxB}$}
	\setlength\mylenA{\the\wd\myboxA}
	\addtolength\mylenA{-\the\wd\myboxB}%
	\ifdim\wd\myboxB<\wd\myboxA%
	\rlap{\hskip 0.5\mylenA\usebox\myboxB}{\usebox\myboxA}%
	\else
	\hskip -0.5\mylenA\rlap{\usebox\myboxA}{\hskip 0.5\mylenA\usebox\myboxB}%
	\fi}
\makeatother
\usepackage[thicklines]{cancel}
\begin{document}
	
\title{Bubble wall velocity during electroweak phase transition in the inert doublet model}
	
\author{Siyu Jiang}

\author{Fa Peng Huang}
\email{Corresponding Author. 
 huangfp8@mail.sysu.edu.cn}

\author{Xiao Wang}%

\affiliation{MOE Key Laboratory of TianQin Mission, TianQin Research Center for
Gravitational Physics \& School of Physics and Astronomy, Frontiers
Science Center for TianQin, Gravitational Wave Research Center of CNSA, 
Sun Yat-sen University (Zhuhai Campus), Zhuhai 519082, China}

\bigskip
	
\date{\today}
	
\begin{abstract}
The bubble wall velocity is essential for the spectra of phase transition gravitational wave, electroweak baryogenesis and the dynamical dark matter mechanism. We perform detailed calculations of the bubble wall velocity in the well-motivated inert doublet model using the microphysical approach with some recent new methods.
\end{abstract}
	
\maketitle

\section{Introduction}
Many fundamental problems, like the electroweak baryogenesis~\cite{Cline:2020jre},  the dark matter (DM) formation mechanism~ \cite{Baker:2019ndr,Chway:2019kft,Huang:2017kzu} and the strength or energy budget~\cite{Espinosa:2010hh,Giese:2020rtr,Giese:2020znk,Wang:2020nzm,Leitao:2010yw,Leitao:2014pda} of the phase transition gravitational wave (GW) strongly depend on the precise value of the bubble wall velocity. For traditional electroweak baryogenesis, bubble wall velocities lower than the sound speed are favored to guarantee sufficient diffusion time for the baryons. The net baryon-to-photon ratio observed by the cosmic microwave background  or big bang nucleosynthesis may be sensitive to the bubble wall velocity. New mechanisms for baryogenesis with large 
bubble wall velocity were recently proposed in Refs.~\cite{Laurent:2020gpg,DeCurtis:2022hlx,Laurent:2022jrs}.
Thus, to understand the origin of the matter-antimatter asymmetry of our Universe, precise calculation of 
the bubble wall velocity is crucial for a given electroweak baryogenesis model.
{Recently}, motivated by the current situation of DM detection,  a new mechanism is
proposed to produce DM by the bubble dynamics through a strong first-order phase transition (SFOPT)
process in the early Universe~\cite{Baker:2019ndr,Chway:2019kft,Huang:2017kzu}. For this new DM mechanism, the bubble wall velocity is
also important as shown in Ref.~\cite{Huang:2017kzu}. The DM relic density might depends on the bubble wall velocity.  Both the electroweak baryogenesis and the new DM mechanisms could be probed by phase
transition GW signals, whose spectra strongly depend on the bubble wall velocity. The GW spectra for subsonic bubble wall velocity and supersonic bubble wall velocity might have several orders of hierarchy.
For example, the energy budget of the phase transition GW is sensitive to the bubble wall velocity.
Therefore, from the perspective of GW experiments, bubble wall velocity would be the first dynamical parameter of a SFOPT to be confirmed if the phase transition GW is observed at future GW experiments like LISA~\cite{LISA:2017pwj},  TianQin~\cite{TianQin:2015yph,TianQin:2020hid}, and Taiji~\cite{Hu:2017mde}.

The bubble wall velocity is very important for both theoretical study and experimental search. However,
the precise calculation of bubble wall velocity is complicated and difficult. 
The bubble wall velocity of a given new physics model is generally determined by  the hydrodynamic and microphysical processes which basically involve the particles scattering processes between the relevant particles of the new model at the vicinity of the bubble wall. Besides the complicated hydrodynamic effects, the bubble wall velocity might be model dependent based on the fact that
it is complicated to quantify these particle scattering processes precisely in the thermal plasma due to the
thermal effects and the infrared behavior. 
These issues make it difficult to precisely predict the bubble wall velocity.
Most of the previous studies on electroweak baryogenesis, phase transition GW, and phase transition related DM mechanism just take the  bubble wall velocity as an input parameter. 
Therefore, there exists large theoretical uncertainties for these predictions due to the model-dependent bubble wall velocity.   
Some pioneering works \cite{Dine:1992wr,Liu:1992tn,Ignatius:1993qn,Moore:1995si,Moore:1995ua,Moore:2000wx} show us how to calculate the wall velocity under certain assumptions. The bubble wall velocity depends on the friction force acting on the expanding bubbles, which is determined by the deviation of the massive particle populations from thermal equilibrium. Basically, under semiclassical approximation,  the bubble wall velocity can be obtained by simultaneously solving the equation of motion for the Higgs field (or the order-parameter scalar field for the phase transition process) and the Boltzmann equations of the massive particle species.  To solve the Boltzmann equations, it is crucial to calculate the collision terms, which quantify the particle scattering processes in the vicinity of bubble wall.

The bubble wall velocity was first calculated microphysically in Refs.~\cite{Moore:1995ua,Moore:1995si} by Moore and Prokopec for the Standard Model (SM) case and then in Minimal Supersymmetric Standard Model~\cite{John:2000zq}. To avoid their expansive calculations, some works modeled the friction by using phenomenological method~\cite{Espinosa:2010hh,Huber:2013kj,Megevand:2013hwa,Megevand:2009ut} and they also considered  the hydrodynamic effects in detail to the velocity calculation. As the method of Moore and Prokopec is problematic when the wall velocity is approaching the speed of sound, Refs.~\cite{Laurent:2020gpg,DeCurtis:2022hlx,Lewicki:2021pgr,Laurent:2022jrs} proposed some new ansatz, and Refs.~\cite{Dorsch:2021ubz,Dorsch:2021nje} did higher-order calculations. Recently, it is also interesting to consider bubble wall velocity in local equilibrium~\cite{BarrosoMancha:2020fay,Balaji:2020yrx,Ai:2021kak,Wang:2022txy}.

\begin{figure}[htbp]
\centering
\includegraphics[scale=0.06]{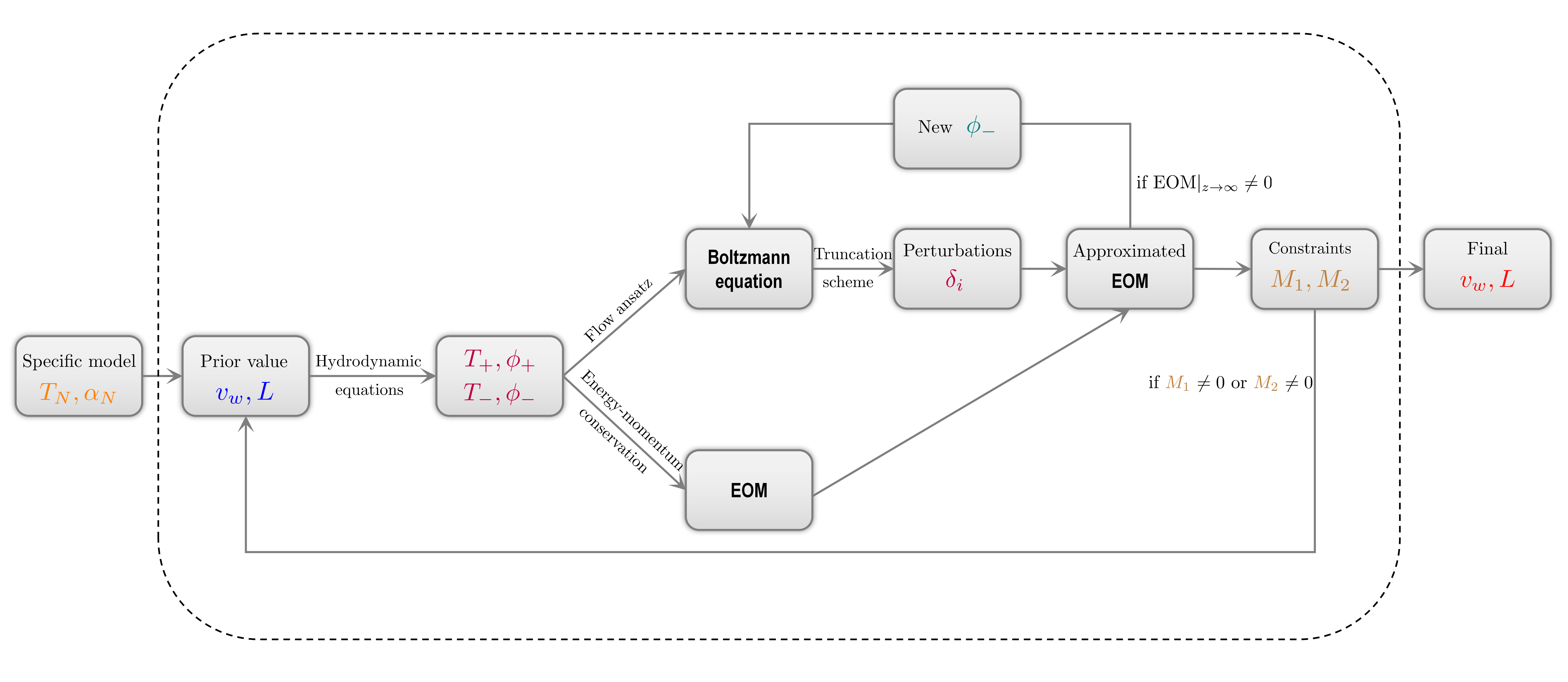}
\caption{ Schematic procedure to calculate the bubble wall velocity during a cosmological phase transition.}\label{process}
\end{figure}

Using the microphysical approach with the recent new methods, for the
first time we calculate the bubble wall velocity in the well-motivated inert doublet model (IDM), which could help to greatly reduce the large uncertainties in 
calculating the phase transition GW and electroweak baryogenesis from the unknown bubble wall velocity. We show the basic procedure to calculate the bubble wall velocity in a given new physics model with a SFOPT in Fig.~\ref{process}.	
	
This paper is organized as follows.
We briefly discuss the IDM and choose the benchmark parameters in Sec.~II,
and describe the equation of motion for the Higgs background field in Sec.~III. 
Then we derive the hydrodynamic effects in the calculation of bubble wall velocity in  Sec.~IV.
In Sec.~V, taking the IDM model as concrete model, we introduce the basic method to 
calculate the bubble wall velocity.
 After that, we calculate the collision terms of the IDM in Sec.~VI. Therefore, we 
obtain the final bubble wall velocity in Sec.~VII.
The conclusion is given in Sec.~VIII.
	
\section{A case study: inert doublet model}
Firstly, we briefly introduce the well-studied IDM, which could improve the naturalness problem and provide a natural DM candidate~\cite{Barbieri:2006dq,LopezHonorez:2010eeh}. Meanwhile, the IDM could trigger a SFOPT~\cite{Chowdhury:2011ga} in the early Universe around 100 GeV during the electroweak spontaneously symmetry breaking process. 
Then, one could use the phase transition GW to explore the IDM. 
However, the GW spectra strongly depend on the bubble wall velocity, which has not been calculated before in the IDM. In this work, we perform detailed calculations on the bubble wall velocity.
We begin our discussions from the tree-level scalar potential at zero temperature,
\begin{align}
V_0 =&\mu^2_1|\Phi|^2+\mu^2_2|\eta|^2
+\frac{1}{2}\lambda_1|\Phi|^4+\frac{1}{2}\lambda_2|\eta|^4  \nonumber\\
&+\lambda_3|\Phi|^2|\eta|^2
+\lambda_4|\Phi^\dagger\eta|^2
+\frac{1}{2}\{ \lambda_5(\Phi^\dagger\eta)^2+\mathrm{H.c.}\}\,\,, \label{eq:potential}
\end{align}
where $\Phi$ is the SM Higgs doublet and $\eta$ represents the inert doublet.
The vacuum stability requires $\lambda_1^{} > 0$, $\lambda_2^{} > 0$, $ \sqrt{\lambda_1^{} + \lambda_2^{}} + \lambda_3^{} > 0$,  $\lambda_3^{}+\lambda_4^{} \pm |\lambda_5^{}| > 0$ ~\cite{Barbieri:2006dq,LopezHonorez:2010eeh}.
At zero temperature, the two doublet scalar fields could be written as
\begin{equation}
\Phi=
\begin{pmatrix}
G^+\\
\frac{1}{\sqrt{2}}(h+v+i G^0)
\end{pmatrix}
,\  \eta=
\begin{pmatrix}
H^+\\
\frac{1}{\sqrt{2}}(H+iA)
\end{pmatrix},
\end{equation}
where the mass of SM Higgs boson $h$ is 125~GeV and the vacuum expectation value (VEV) $v=246$~GeV.
$G^{\pm}$ and $G^0$ are the charged and neutral Nambu-Goldstone bosons respectively. $H$ and $A$ are the $CP$-even and $CP$-odd inert scalars respectively. $H^{\pm}$ are the charged inert scalars.
At zero temperature, we show the inert scalar masses beyond the SM
\begin{align}
\xoverline m_{H}^2&=\mu_2^2+\frac{1}{2}(\lambda_3+\lambda_4+\lambda_5)v^2, \\
\xoverline m_{A}^2&=\mu_2^2+\frac{1}{2}(\lambda_3+\lambda_4-\lambda_5)v^2, \\
\xoverline m^2_{H^{\pm}}&=\mu_2^2+\frac{1}{2}\lambda_3 v^2 \,\,.
\end{align}
These new inert scalars could modify $T$ parameter with the deviation $\Delta T$ approximated as 
\begin{align}\label{tp}
\Delta T \simeq \frac{1}{6\pi e^2 v^2}(\xoverline m_{H^{\pm}}-\xoverline m_{H})(\xoverline m_{H^{\pm}}-\xoverline m_{A})\,\,.
\end{align}
From the above expression, there exists a simple and obvious parameter region that $\Delta T \simeq 0$ if we assume 
$\xoverline m_{A}^2=\xoverline m^2_{H^\pm}$.
This assumption corresponds to $\lambda_4=\lambda_5<0,~~\lambda_3>0$
, and could satisfy all the constraints from electroweak precise measurements, DM direct searches and the collider data~\cite{Chowdhury:2011ga}.
In other words, we assume degenerate  $CP$-odd and  charged scalar masses
\begin{equation}\label{Amass}
\xoverline m_{A}^2=\xoverline m^2_{H^\pm}=\mu_2^2+\frac{1}{2}\lambda_3 v^2   \,\,.
\end{equation}
This makes $\xoverline m_{H}^2=\mu_2^2+\lambda_{L}v^2$ the lightest particle, which becomes the natural DM candidate~\cite{Barbieri:2006dq,LopezHonorez:2010eeh} with
the tiny DM-Higgs boson coupling  $\lambda_{L}=(\lambda_3+\lambda_4+\lambda_5)/2$.
It should be very small to be consistent with the DM direct search. 
Using the package micrOMEGAs~\cite{Barducci:2016pcb}, we include the resonant effects and require the DM relic abundance~\cite{Planck:2018vyg} $\Omega_{\text{DM}} h^2 =  0.11933 \pm 0.00091$.  Then we could obtain the constraint~\cite{Wang:2020wrk}  $\lambda_L \lesssim 0.003$.
These constraints almost reach the blind spots of the IDM, which are difficult for DM direct searches. 
Future lepton colliders in synergy with phase transition GW might help to explore the DM blind spots~\cite{Huang:2017rzf,Wang:2020wrk}.
When $\lambda_3$, $\lambda_4$, $\lambda_5$ are $\mathcal{O}(1)$, a SFOPT with associated phase transition GW could be produced~\cite{Chowdhury:2011ga,Borah:2012pu, Gil:2012ya,Cline:2013bln,AbdusSalam:2013eya,
Blinov:2015vma,Cao:2017oez,Huang:2017rzf,
Laine:2017hdk,Senaha:2018xek,Huang:2019riv,
Kainulainen:2019kyp,Wang:2020wrk}.  

When we discuss the phase transition dynamics of the SFOPT, we denote the background of the Higgs field as $\phi$, namely, 
\begin{equation}
\Phi=
\begin{pmatrix}
0\\
\frac{1}{\sqrt{2}} \phi
\end{pmatrix}\,\,.
\end{equation}
Then we could have the following new field-dependent masses beyond the SM,
\begin{align}
	m_H^2(\phi)&=\mu_2^2+\frac{1}{2}(\lambda_3+\lambda_4+\lambda_5)\phi^2, \\
	m_A^2(\phi)&=\mu_2^2+\frac{1}{2}(\lambda_3+\lambda_4-\lambda_5)\phi^2, \label{mA} \\
	m^2_{H^{\pm}}(\phi)&=\mu_2^2+\frac{1}{2}\lambda_3 \phi^2 \label{mH} \,\,.
\end{align}
We should obtain the finite-temperature effective potential~\cite{Niemi:2021qvp,Schicho:2022wty} after taking the one-loop quantum and thermal corrections with daisy resummation into account. Namely, 
\begin{equation}
V_{\text{eff}}(\phi, T)=V_0(\phi)+V_{\text{CW}}({\phi})+V_{\text{T}}(\phi, T)\,\,.
\end{equation}
The one-loop quantum correction of the potential in the on-shell renormalizaiton scheme is
\begin{equation}
V_{\text{CW}}({\phi}, T=0)=\sum_i \frac{n_i}{64 \pi^2}\left[{m}_i^4(\phi)\left(\ln \frac{{m}_i^2(\phi)}{\xoverline m_i^2}-\frac{3}{2}\right)+2 \xoverline m_i^2 \bar{m}_i^2(\phi)\right],
\end{equation}
where $n_i$ is the degree of freedom for each massive particle $m_i(\phi)$.
For the daisy resummation, we use the scheme proposed by Dolan and Jackiw~\cite{Dolan:1973qd} where one only needs to consider the thermal correction in the thermal function $I_b$. Namely, the one-loop thermal correction $V_{\text{T}}$  should be improved as
\begin{equation}
V_{\text{T}}(\phi, T>0)=\sum_i n_i \frac{T^4}{2 \pi^2} I_b \left(\frac{{M}_i^2}{T^2}\right),
\end{equation}
where ${M}_i^2={m}_i^2(\phi)+\Pi_i(T)$ and
$I_{b, f}\left(a^2\right)=\int_0^{\infty} d x ~x^2 \ln \left(1 \mp e^{-\sqrt{x^2+a^2}}\right)$.
$\Pi_i(T)$ represents the thermal correction for scalar bosons and the
longitudinal components of gauge bosons.
Scalars with
$\mu_1^2$ and $\mu_2^2$ appearing in their field-dependent masses can be simply replaced by $\mu_1^2+\Pi_{\Phi}(T)$ and $\mu_2^2+\Pi_\eta(T)$, respectively, where
\begin{align}
	\Pi_\Phi &= \frac{T^2}{12}\left[3\lambda_1 + 2\lambda_3 + \lambda_4 + \frac{3}{4}(3g_w^2 + g_Y^2) + 3y_t^2\right],\\
	\Pi_\eta &= \frac{T^2}{12}\left[3\lambda_2 + 2\lambda_3 + \lambda_4 + \frac{3}{4}(3g_w^2 + g_Y^2)\right].
\end{align}
For the gauge bosons, only their longitudinal components need to include the thermal corrections in the thermal function $I_b$ as the following: 
\begin{equation}
\begin{aligned}
	&\Pi_W(T)=\Pi_W^{\Phi}(T)+\Pi_W^{\eta}(T)=\left[\frac{11}{6}+\frac{1}{6}\right] g_w^2 T^2=2 g_w^2 T^2 \,\,,\\
	&\Pi_B(T)=\Pi_B^{\Phi}(T)+\Pi_B^{\eta}(T)=\left[\frac{11}{6}+\frac{1}{6}\right] g_Y^2 T^2=2 g_Y^2 T^2 \,\,,
\end{aligned}
\end{equation}
where $g_w$ and $g_Y$ are the gauge coupling of $SU(2)_L$ and $U(1)_Y$, respectively. Hence, 
for the longitudinal components of the gauge bosons, their physical masses  are eigenvalues of the following matrix
\begin{equation}
	{M}_{\text{L}}^2 = \left(\begin{array}{cccc}
		m_1^2 + \Pi_W & 0 & 0 & 0\\
		0 & m_1^2 + \Pi_W & 0 & 0\\
		0 & 0 & m_1^2 + \Pi_W & m_{12}^2\\
		0 & 0 & m_{12}^2 & m_2^2 + \Pi_B\\
	\end{array}\right) \,\,,
\end{equation}
where $m_1^2=g_w^2 \phi^2/4$, $m_2^2=g_Y^2 \phi^2/4$ and $m_{12}^2=-g_w g_Y \phi^2/4$.

 To focus on the calculation of bubble wall velocity during the SFOPT in the IDM, we simply choose three sets of benchmark model parameters based on our numerical calculations where
we use the package micrOMEGAs~\cite{Barducci:2016pcb} for DM relic abundance, DM direct search, collider constraints~\cite{Belyaev:2016lok}, and use 
CosmoTransitions~\cite{Wainwright:2011kj} to calculate the phase transition dynamics.
Taking all the above discussions into consideration, we choose the following benchmark parameters where
$\xoverline m_h=125$~GeV,
$\lambda_2=0.2$. Other parameters are shown in Table~\ref{ptable}.
For these benchmark point sets, the DM relic density, the DM direct search, the collider constraints and condition of the SFOPT could be satisfied simultaneously.
We perform detailed calculations of the bubble wall velocity based on Benchmark A and give the final results of bubble wall velocity for Benchmarks B and C.  

\begin{table}[t]
	\centering
	
	\setlength{\tabcolsep}{3mm}
	
	\begin{tabular}{c|c|c|c|c|c}
		\hline\hline
		&$\xoverline{m}_H$~[GeV] & $\xoverline{m}_A=\xoverline{m}_{H^{\pm}}$~[GeV]&$\lambda_L$ &$T_c $~[GeV] &$T_N$~[GeV]\\
		\hline
	Benchmark A	&62.66 & 300 &0.0015 &118.3 &117.1  \\
		\hline
	Benchmark B	&65.00 &300 &0.0015 &118.6 &117.5  \\
		\hline
	Benchmark C	&63.00 & 295 &0.0015 &119.4 & 118.4  \\
		\hline\hline
	\end{tabular}
	\caption{Three sets of benchmark model parameters.}\label{ptable}
\end{table}

\section{Standard method for solving bubble wall velocity}

After bubbles of the broken phase are nucleated, they will expand at a steady bubble wall velocity under some certain circumstances.
To obtain the reliable bubble wall velocity in a given new physics model,
we firstly need to know the bubble dynamics which are described by the equation of motion (EOM) of the background field for the SFOPT process. 
We can obtain the EOM by using energy-momentum conservation of the scalar-plasma system. The energy-momentum tensor of the field is
\begin{equation}
	T^{\mu \nu}_\phi=\partial^\mu \phi \partial^\nu \phi-g^{\mu \nu}\left(\frac{1}{2} \partial_\alpha \phi \partial^\alpha \phi-V_{T=0}(\phi)\right) \,\,,
\end{equation}
where $\phi$ is the background Higgs field in the IDM, $V_{T=0}(\phi)=V_0(\phi)+V_{\rm CW}(\phi)$ is the effective potential at zero temperature.
The energy-momentum tensor of the plasma is
\begin{equation}
	T_{\mathrm{pl}}^{\mu \nu}=\sum_{i} \int \frac{\mathrm{d}^3 {p}}{(2 \pi)^3} \frac{p^\mu p^\nu}{E_i} f_i(x, p) \,\,,
\end{equation}
where $f_i(x,p)$ is the distribution function of the particle and $E_i=\sqrt{p^2+m_i^2}$. The sum is over all particle species.
Using energy-momentum conservation condition
\begin{equation}
	\nabla_{\mu}(T_{\phi}^{\mu \nu}+T_{\rm {pl}}^{\mu \nu})=0\,\,,
\end{equation}
we can derive the EOM of the background  field~\cite{Moore:1995ua,Moore:1995si,Espinosa:2010hh,Konstandin:2014zta}, 
\begin{equation}\label{eom0}
	\Box\phi + \frac{\partial V_{T=0}(\phi)}{\partial \phi} + \sum_i\frac{dm_i^2}{d\phi}\int\frac{d^3p}{(2\pi)^32E_i} f_i(x,p) = 0 \,\,,
\end{equation}
During a SFOPT, some massive particles will be in an out-of-equilibrium state near the bubble wall.
Hence the distribution function could be approximated as a thermal equilibrium part plus an out-of-equilibrium part, namely, 
 $$f_i \equiv f_{0,i} + \delta f_i \,\,.$$
The equilibrium distribution function for fermions and bosons in the plasma frame are given by $f_{0,i} = \frac{1}{\exp(E_i/T)\pm1}$, respectively.
The integral of the equilibrium part of the distribution functions gives the thermal-correction part of the effective potential $V_{\rm T}(\phi,\text{T})$.
Therefore, the EOM for the background field could be further simplified to 
\begin{equation}\label{eomt}
	\Box\phi + \frac{\partial V_{\rm eff}(\phi,T)}{\partial\phi} + \underbrace{\sum_i \frac{dm_i^2}{d\phi}\int\frac{d^3p}{(2\pi)^32E_i}\delta f_i(x,p) }_{\mathclap{\text{friction term}}}= 0\,\,,
\end{equation}
where $V_{\text{eff}}(\phi, T)=V_0(\phi)+V_{\text{CW}}({\phi})+V_{\text{T}}(\phi,T)$ is the thermal effective potential in a specific model. The second term is the driving term that accelerates the bubble wall.
The third term in Eq.~\eqref{eomt} behaves as the friction force acting on the bubble wall. We can see that the contributions mainly come from ``heavy'' particles with non-negligible field-dependent mass. ``Heavy'' means that the particle could get field-dependent mass comparable with the temperature. Thus, in the IDM, we could only consider the contributions from $t,W^{\pm},Z,A,H^{\pm}$. For the stationary wall, choosing the $-z$ as the propagating direction as shown in Fig.~\ref{wallshock},  all quantities $Q$ are functions of $z+v_wt$, so we can set $z+v_w t\rightarrow z$ and hence $\partial_tQ\rightarrow v_w Q'$, $\partial_zQ\rightarrow Q'$. $v_w$ is the bubble wall velocity and prime means derivative with respect to $z$. Then in the \emph{plasma frame}, the EOM with bubble wall velocity could be further expressed as in Refs.~\cite{Moore:1995si,Moore:1995ua}
\begin{equation}\label{eom3}
	(1-v_w^2)\phi'' + \frac{\partial  V_{\rm eff}(\phi,T)}{\partial\phi} + \sum_i \frac{dm_i^2}{d\phi}\int\frac{d^3p}{(2\pi)^32E_i}\delta f_i(x,p) = 0 \,\,.
\end{equation}

\section{Hydrodynamic treatment}\label{ht}

\begin{figure}[htbp]
	\centering
	\includegraphics[scale=0.1]{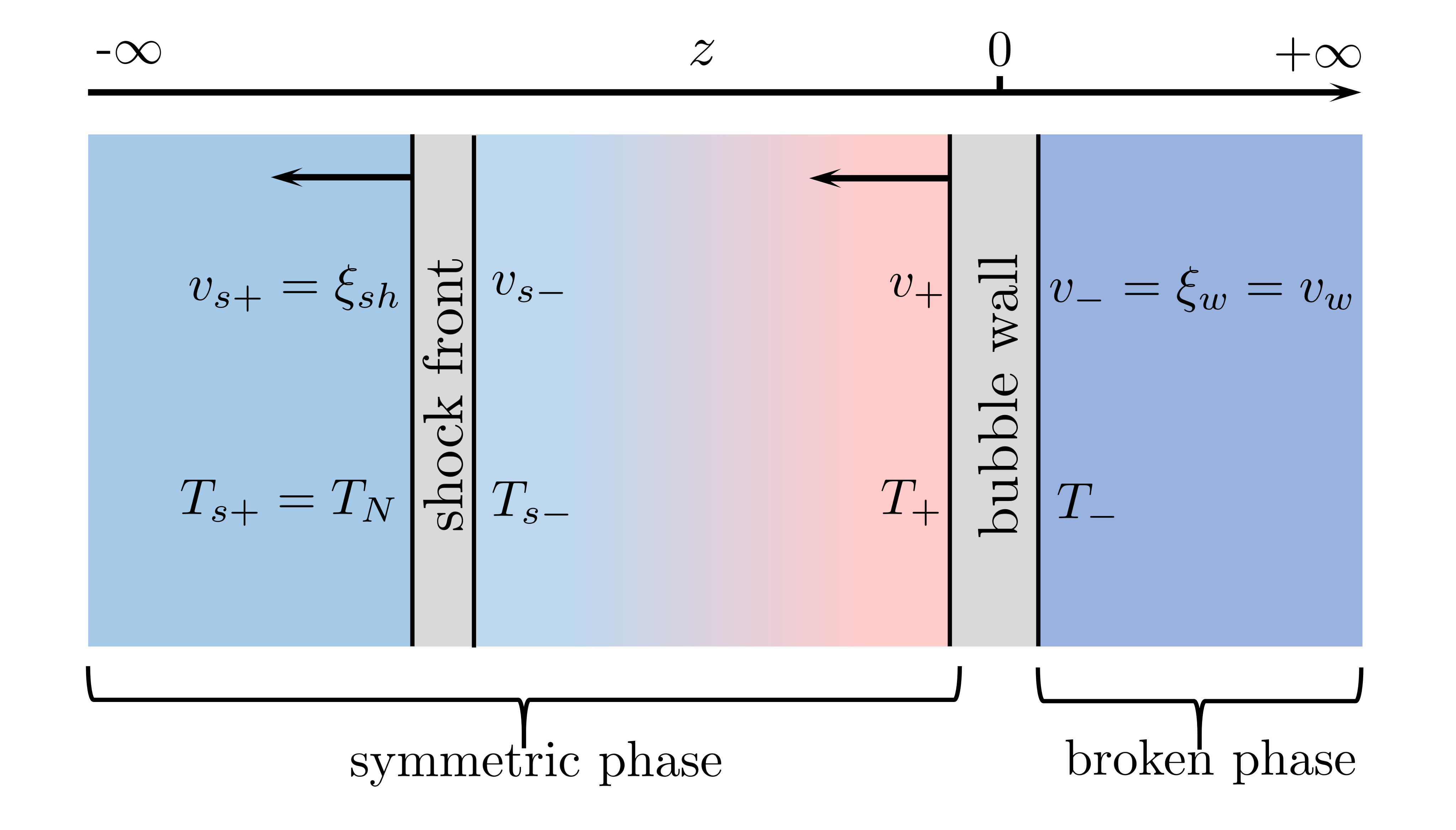}
    \caption{Dynamical process for a bubble wall in the deflagration mode.  The deflagration wall (w) is moving along the negative $z$ direction (as indicated by arrows) with a shock front (s) propagating in front of it. $v_+,T_+$ are the velocity and temperature in front of the wall, and $v_-,T_-$ are the velocity and temperature behind the wall. All $v_-,T_-,v_+,T_+$ are depicted in the bubble wall frame. $v_{s+},T_{s+}$ are the velocity and temperature in front of the shock front and $v_{s-},T_{s-}$ the velocity and temperature behind the shock front. All $v_{s-},T_{s-},v_{s+},T_{s+}$ are depicted in the shock frame.}\label{wallshock}
\end{figure}

Having the above EOM, we begin to consider the hydrodynamic effects where the temperature might play important roles. For example, Eq.~\eqref{eom3} should be evaluated at a specific temperature. Then we should consider the hydrodynamic effects which cause the temperature and velocity profiles varying across the bubble wall~\cite{Wang:2020nzm,Espinosa:2010hh,Leitao:2010yw,Leitao:2014pda,Giese:2020rtr,Giese:2020znk}. This could be seen from the second term of Eq.~\eqref{eom3} in the wall frame where we multiply it by $\phi'$ and integrate over $z$,
\begin{eqnarray}
		\int dz \phi' \frac{\partial V_{\rm eff}(\phi,T)}{\partial \phi} &=&\int dz \frac{dV_{\rm eff}(\phi,T)}{d z}-\int dz\frac{\partial V_{\rm eff}(\phi,T)}{\partial T}\frac{dT}{dz}\\
		&\simeq& V_{\rm eff}(\phi_-,T_-)-V_{\rm eff}(\phi_+,T_+)+\langle s \rangle (T_- -T_+)\,\,, \label{Vs}
\end{eqnarray}
where $V_{\rm eff}(\phi_-,T_-)-V_{\rm eff}(\phi_+,T_+)$ is the pressure difference that push the bubble wall out. Notice that $\phi_-=\phi(z>0)$ and $T_-=T(z>0)$ because we let wall move along the negative $z$ direction as shown in Fig.~\ref{wallshock} . Entropy density $s(\phi,T) \equiv -\partial V_{\rm eff}(\phi,T)/\partial T$ and it gives non-negligible contribution to the pressure. $\langle s \rangle \simeq \frac{s(\phi_+,T_+)+s(\phi_-,T_-)}{2}$ is the average entropy density. Notice that the presence of $T_+$ here in stead of $T_N$ is due to the hydrodynamic heating effects in front of the bubble wall. Besides, as we consider the hydrodynamic effects, temperature in front of the bubble wall $T_{+}$ is usually different from the one behind the wall $T_{-}$. Therefore, it is important to calculate the temperature profile around the bubble wall. All the perturbations are evaluated in front of the wall where $T=T_+$.

The energy-momentum tensor of the equilibrium part of plasma and scalar field can be combined in the form of ideal fluid,
\begin{equation}
	T_{f}^{\mu \nu}=(e_{f}+p_{f})u^{\mu}u^{\nu}+p_{f}g^{\mu \nu}=\omega_{f}u^{\mu}u^{\nu}+p_{f}g^{\mu \nu}
\end{equation} 
where $e_{f}$ is the energy density, $p_{f}$ is the pressure and $\omega_{f}$ is the enthalpy of the fluid. $u^\mu$ is the fluid four-velocity $u^\mu=\gamma(v)(1, \vec{v})$ in the background plasma frame with the Lorentz factor $\gamma(v)=1 / \sqrt{1-v^2}$. 

We can write the hydrodynamic equations in the rest frame of Universe \cite{Espinosa:2010hh}, 
\begin{eqnarray}
	\begin{gathered}
		(\xi-v) \frac{\partial_{\xi} e_{f}}{\omega_f}=2 \frac{v}{\xi}+\left[1-\gamma^2 v(\xi-v)\right] \partial_{\xi} v,\\
		(1-v \xi) \frac{\partial_{\xi} p_{f}}{\omega_f}=\gamma^2(\xi-v) \partial_{\xi} v\,\,,
	\end{gathered}\label{hydro}
\end{eqnarray}
where $\xi=r / t$ is the self-similar variable of the above equations and has unit of velocity. $r$ is the distance to the bubble center and
$ t$ is the time since the bubble nucleation. ``Self-similar" means that there is no characteristic length or time scale for one steadily expanding bubble.  $v\left(\xi_w\right)$ is the fluid velocity at the location of the bubble wall $\xi_w$. Notice that $\xi_w=v_w$. 

We can obtain the differential equation for the temperature from Eq.~\eqref{hydro} by using the following identity
\begin{equation}
	\frac{\partial p_{f}}{\partial T}=\partial_{\xi} p_{f} \frac{\partial \xi}{\partial T}\,\,,
\end{equation}
then the enthalpy can be written as
\begin{equation}
	\omega_{f} \equiv T \frac{\partial p_{f}}{\partial T}=T \partial_{\xi} p_{f}\left(\partial_{\xi} T\right)^{-1}\,\,.
\end{equation}
Substituting it into Eq.~\eqref{hydro} one obtains
\begin{equation}\label{tpro}
	\frac{\partial_{\xi} T}{T}=\gamma^2 \mu \partial_{\xi} v \,\,,
\end{equation}
where $\mu(\xi, v)=(\xi-v)/(1-\xi v)$ is Lorentz transformation of the velocity.
Given speed of sound in the plasma, $c_s^2 \equiv(d p_{f} / d T) /(d e_{f} / d T)$, from Eq.~\eqref{hydro} we can get the equation of the velocity profile,
\begin{equation}\label{vpro}
	2 \frac{v}{\xi}=\gamma^2(1-v \xi)\left[\frac{\mu^2}{c_s^2}-1\right] \partial_{\xi} v \,\,.
\end{equation}

In order to solve the hydrodynamic fluid equations, one needs to consider the boundary conditions in the system by using the conservation of the energy momentum tensor $\nabla_{\mu} (T_{\phi}^{\mu \nu}+T_{\rm pl}^{\mu \nu})=\nabla_\mu T_{f}^{\mu\nu}=0$ across the bubble wall,
\begin{equation}
	\omega_{+}\gamma_{+}^2 v_{+}^2+p_+=\omega_{-}\gamma_{-}^2 v_{-}^2+p_-, \quad \omega_{+}\gamma_{+}^2 v_{+}=\omega_{-}\gamma_{-}^2 v_{-}\,\,, \label{match}
\end{equation}
which can be transformed as
\begin{equation}
	v_{+} v_{-}=\frac{p_{+}-p_{-}}{e_{+}-e_{-}}, \quad \frac{v_{+}}{v_{-}}=\frac{e_{-}+p_{+}}{e_{+}+p_{-}}\,\,,
\end{equation}
where $+(-)$ means in front of (behind) the bubble wall. These quantities can be estimated by using the so-called bag model. In the bag model, the pressure and the energy density can be written as
\begin{eqnarray}
\begin{aligned}
	&p_{+}=\frac{1}{3} a_{+} T_+^4-\epsilon_{+}, \quad e_{+}=a_{+} T_+^4+\epsilon_{+}, \quad \text{where} \quad \epsilon_{+} \equiv V_{T=0}\left(\phi_{+}\right) ; \\
	&p_{-}=\frac{1}{3} a_{-} T_-^4-\epsilon_{-}, \quad e_{-}=a_{-} T_-^4+\epsilon_{-}, \quad \text{where} \quad \epsilon_{-} \equiv V_{T=0}\left(\phi_{-}\right) \text {. } 
\end{aligned}
\end{eqnarray}
Notice that these conditions are derived in the rest frame of the bubble wall. We have pointed out that in order to solve the bubble wall velocity, one important quantity is the temperature in front of the wall $T_+$, where massive particles meet the bubble wall. Actually, $v_{+},T_{+}$ are important not only for solving wall velocity but also for calculating electroweak baryogenesis. In bag model we can see the first two terms in Eq.~\eqref{Vs},
\begin{eqnarray}
	V_{\rm eff}(\phi_-,T_-)-V_{\rm eff}(\phi_+,T_+)&=&p_+-p_-=\epsilon_--\epsilon_+ +\frac{1}{3} a_{+} T_+^4-\frac{1}{3} a_{-} T_-^4 \notag \\
	&=&\epsilon_--\epsilon_+ +\frac{1}{3}a_-(T_+^4-T_-^4) +\frac{1}{3}(a_+-a_-)T_+^4 \,\,.
\end{eqnarray}
For weak phase transitions $T_+ \simeq T_-$ the third term is approximately zero and the fourth term accounts for entropy variation which may behave as friction in \emph{local equilibrium} as discussed in Refs.~\cite{Konstandin:2010dm,BarrosoMancha:2020fay,Balaji:2020yrx,Ai:2021kak,Wang:2022txy}.

Having assumed that the system could be approximated by the bag equation of state, one could obtain a relation between the plasma velocities in front of and behind the bubble wall,
\begin{equation}
	v_{+}=\frac{1}{1+\alpha_+}\left[\left(\frac{v_{-}}{2}+\frac{1}{6 v_{-}}\right) \pm \sqrt{\left(\frac{v_{-}}{2}+\frac{1}{6 v_{-}}\right)^2+\alpha_+^2+\frac{2}{3} \alpha_+-\frac{1}{3}}\right] \label{vpm} \,\,,
\end{equation}
where $\alpha_+=(\epsilon_+-\epsilon_-)/(a_+T_+^4)$ is the phase transition strength. There would be three possible stable solutions to the hydrodynamic equations~\cite{Steinhardt:1981ct}: 
1) Deflagration solution corresponding to the minus sign of above formula has a subsonic wall velocity, and $v_+<v_-$. In order to fulfill the requirement that the velocity far away from the wall should be zero, there should be a shock front in front of the bubble wall, as we can see in Fig.~\ref{wallshock}. Note that in this case, the fluid behind the wall is at rest so we have $v_{-}=v_w$; 
2) For the detonation solution,  the wall velocity is supersonic and the fluid in front of the wall is at rest. In this case the solution corresponds to the positive sign of Eq.~\eqref{vpm} and we have $v_+=v_w$; 
3)The hybrid expansion occurs when $v_+<v_-=c_s$. This is a hybrid state that there should be a shock front in front of the wall and a rarefaction wave behind the wall.

In this work, we concentrate on deflagration mode where a shock front is formed in front of the bubble wall as shown in Fig.~\ref{wallshock}. We can clearly see that there exist two different interface boundaries, which lead to discontinuous thermodynamic quantities and two different reference frames. Therefore, we should carefully mention the designated reference frame when we deal with the velocity. For convenience, 
we denote the fluid velocities as measured with respect to the wall frame as $v_{+}$ if the fluid is in front of the wall and as $v_{-}$ if the fluid is behind the wall. And we denote the fluid velocities as measured with respect to the shock frame as $v_{s+}$ if the fluid is in front of the shock front and as $v_{s-}$ if the fluid is behind the shock front.

For the shock front, the fluid on both sides of it is in the symmetric phase and it has the same matching conditions as bubble wall in Eq.~\eqref{match}, then the velocities on both sides can be written as
\begin{equation}
	v_{s+} v_{s-}=\frac{1}{3}, \quad \frac{v_{s+}}{v_{s-}}=\frac{T_{s+}^4+3 T_{s-}^4}{T_{s-}^4+3 T_{s+}^4} \,\,. \label{shock}
\end{equation}
We use tildes $\tilde{v}_{\pm}$ to refer to fluid velocities in the fluid reference frame. This is also the frame of the universe or the reference frame of the center of the bubble. In this frame the fluid far behind the bubble wall and far ahead of the shock front is at rest.
The fluid velocities in wall frame or shock frame and fluid frame are related by Lorentz transformations, i.e.,
\begin{equation}
	\tilde{v}_{\pm}=\frac{v_w-v_{\pm}}{1-v_w v_{\pm}}\,\,, \quad \tilde{v}_{s \pm}=\frac{v_{s h}-v_{s \pm}}{1-v_\omega v_{s \pm}}\,\,,
\end{equation}
where $v_w$ is just the bubble wall velocity, i.e., the relative velocities between the wall and the center of the bubble, $v_{s h}$ is the relative velocity between the shock front and the center of the bubble.

\begin{figure}[htbp]
	\centering
	\subfigure{
		\begin{minipage}[t]{0.5\linewidth}
			\centering
			\includegraphics[scale=0.3]{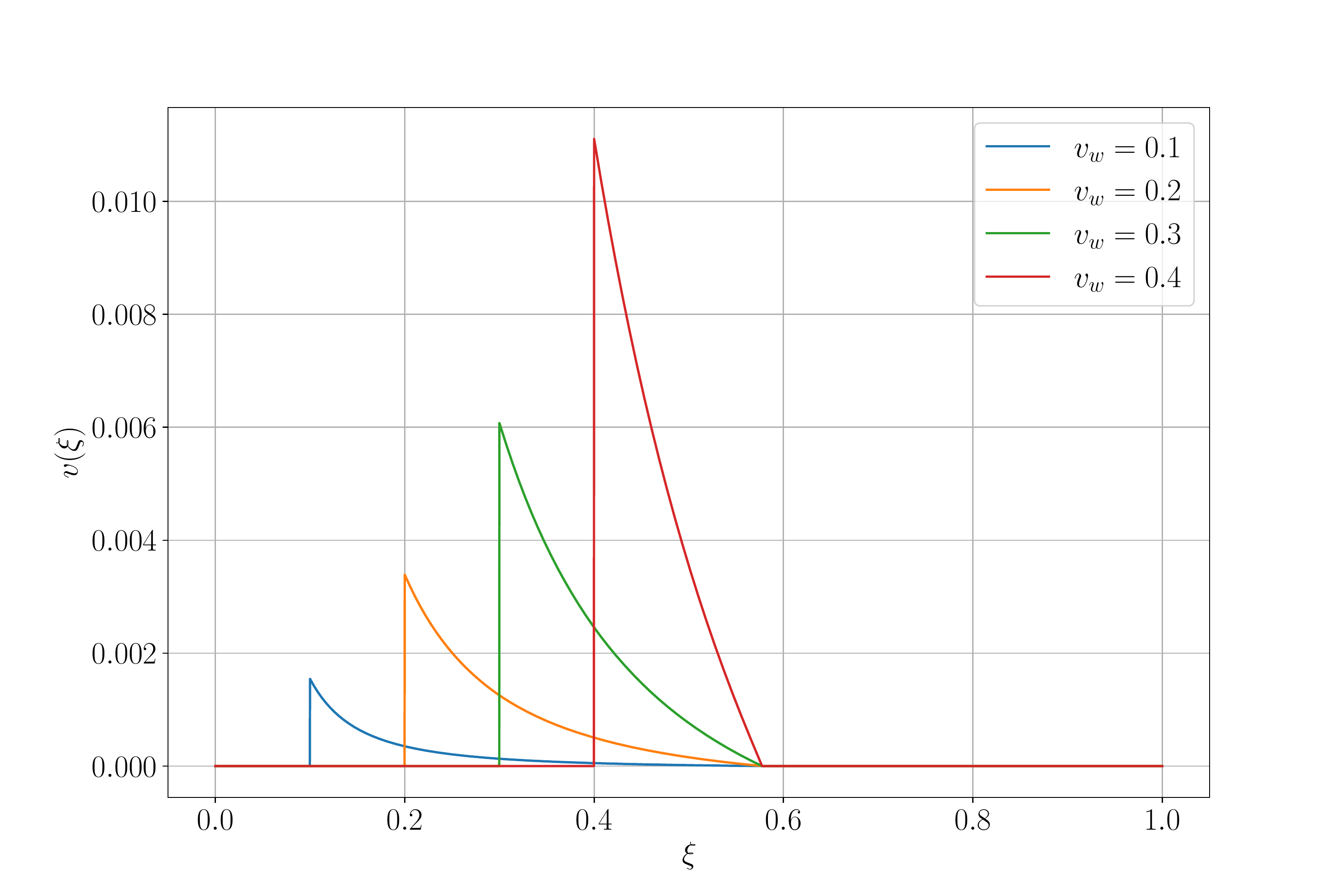}
	\end{minipage}}%
	\subfigure{
		\begin{minipage}[t]{0.5\linewidth}
			\centering
			\includegraphics[scale=0.3]{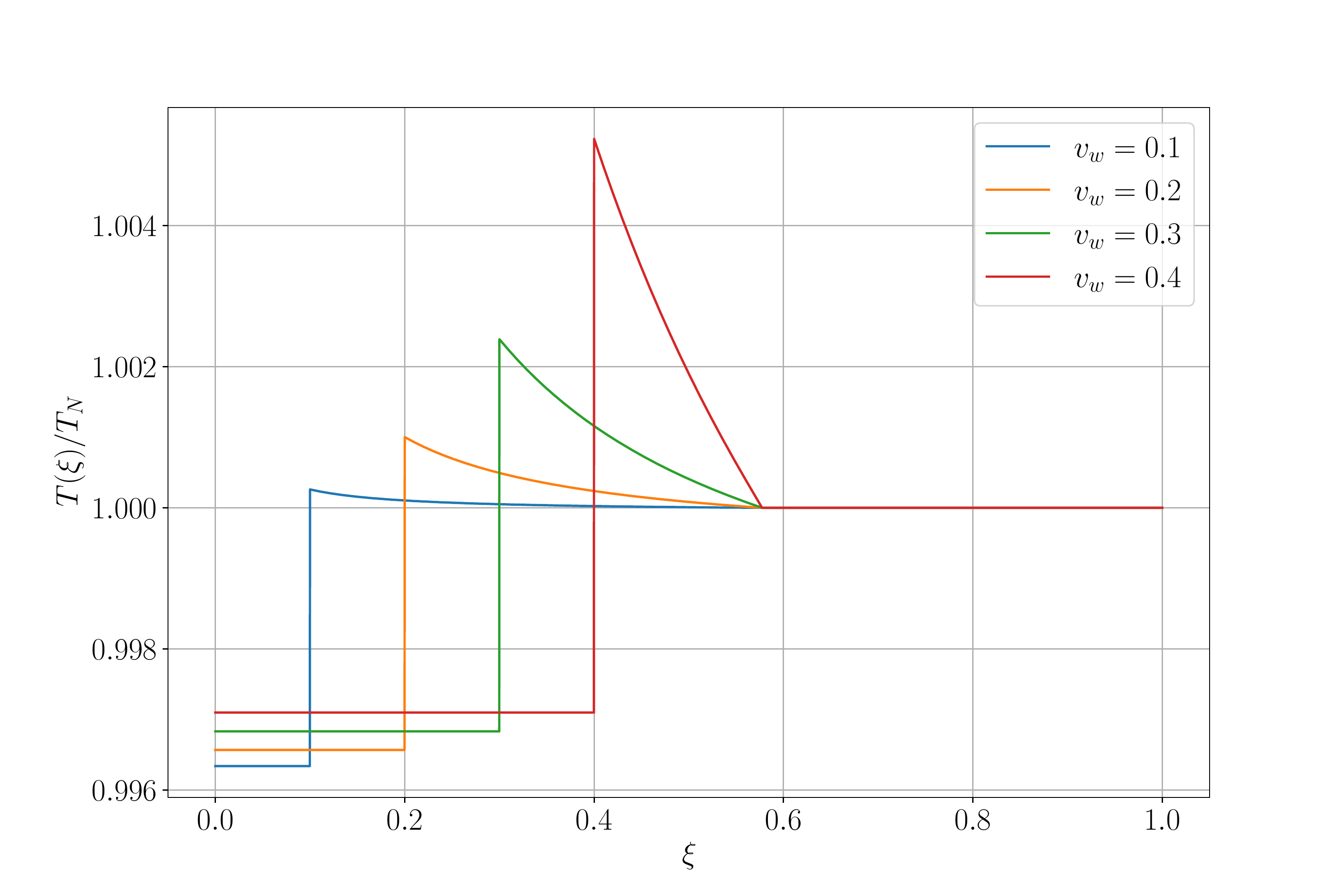}
	\end{minipage}}	
	\caption{Velocity and temperature profiles in the deflagration mode for different bubble wall velocity $v_w=0.1,0.2,0.3,0.4$. Here we choose $\alpha_N=0.005$. Left panel is the velocity profile and right panel is the temperature profile.}\label{vT}
\end{figure}

Since we focus on the subsonic bubble wall velocity, we should give a more precise definition of a deflagration. In this case the fluid velocity behind the wall is zero in the fluid frame, i.e,
\begin{equation}
	\tilde{v}_{\boldsymbol{-}}=0 \quad \longrightarrow \quad v_{w}=v_{\boldsymbol{-}} \,\,.
\end{equation}
Similarly, we require that the fluid far away should be at rest as in the presence of the shock front. Then we will get
\begin{equation}
	\tilde{v}_{s+}=0 \quad \longrightarrow \quad v_{s h}=v_{s+} \,\,.
\end{equation}
In order to obtain the fluid velocity and temperature profile in front of the bubble wall, we need to solve their corresponding hydrodynamic equations \eqref{tpro} and \eqref{vpro}.
Given $\alpha_+$ and some value for the wall velocity $v_w<c_s$, noting we have $\xi_w=v_w=v_{-}$, the boundary condition for the velocity equation in front of bubble wall is
$$
\tilde{v}_{+} \equiv \frac{\xi_w-v_{+}}{1-\xi_w v_{+}}=v\left(\xi_w\right),
$$
with $v_{+}$ which is determined by Eq.~\eqref{vpm}. From Eq.~\eqref{vpro}, we can see that the solution would have a singularity when $\mu(\xi,v)=c_s$. 
However, the solution will reach the shock front position $\xi_{sh}$ before the singularity. The shock front position satisfies the second boundary condition,
\begin{equation}\label{vsml}
	\tilde{v}_{s-} \equiv \frac{\xi_{sh}-v_{s-}}{1-\xi_{sh} v_{s-}}=v\left(\xi_{sh}\right) \,\,.
\end{equation}
Notice that $\xi_{sh}=v_{s h}=v_{s+}$. However, we still do not know the position of the shock front $\xi_{sh}$. One can use the inverse transformation of Eq.~\eqref{vsml} and then substitute it into Eq.~\eqref{shock} which describes the discontinuities of velocities across the shock front, then we will get
\begin{equation}
	\mu\left(\xi_{sh}, v\left(\xi_{sh}\right)\right) \xi_{sh}=\frac{1}{3}=c_s^2 \,\,,
\end{equation}
which determines the position of the shock front. 
Having the velocity profile, we can substitute it into Eq.~\eqref{tpro} to get the temperature profile. The boundary values are the values of the temperature outside the wall $T_{+}$ and the temperature inside the shock front $T_{s-}$. In Fig.~\ref{vT} we show an illustration of velocity profile and temperature profile where we choose $\alpha_N=0.005$. We could integrate Eq.~\eqref{tpro} and get
\begin{equation}\label{TpTsm}
	\frac{T_{+}}{T_{s-}}=\exp \left[\int_{\xi_{sh}}^{\xi_{w}} d \xi \gamma^2 \mu \partial_{\xi} v\right]=\exp \left[\int_{\xi_{sh}}^{\xi_{w }} d \xi \frac{2 c_s^2 v(\xi-v)}{\xi\left((\xi-v)^2-c_s^2(1-v \xi)^2\right)}\right] \,\,.
\end{equation}
Notice that the temperature in front of the shock front corresponds to the nucleation temperature of the transition, i.e., $T_{s+}=T_N$.
Using Eq.~\eqref{shock} and eliminate $v_{s+}$ we can find
\begin{equation}\label{TsmTn}
	\frac{T_{s-}^4}{T_N^4}=	\frac{T_{s-}^4}{T_{s+}^4}=\frac{3\left(1-v_{s-}^2\right)}{9 v_{s-}^2-1}\,\,.
\end{equation}
Using Eqs.~\eqref{TpTsm} and \eqref{TsmTn} the temperature in front of the bubble wall $T_+$ is related to the nucleation temperature $T_{N}$ as 
\begin{equation}
	\frac{T_{+}}{T_N}=\left(\frac{3\left(1-v_{s-}^2\right)}{9 v_{s-}^2-1}\right)^{1 / 4} \exp \left[\int_{\xi_{sh}}^{\xi_{w }} d \xi \frac{2 c_s^2 v(\xi-v)}{\xi\left((\xi-v)^2-c_s^2(1-v \xi)^2\right)}\right] \,\,.
\end{equation}
Then we can use 
\begin{equation}
	\frac{a_+ T_+^4}{a_-T_-^4}=\frac{\omega_+}{\omega_-}\,\,,
\end{equation}
to determine the temperature behind the wall $T_-$.

\begin{figure}[htbp]
	\centering
    \includegraphics[scale=0.43]{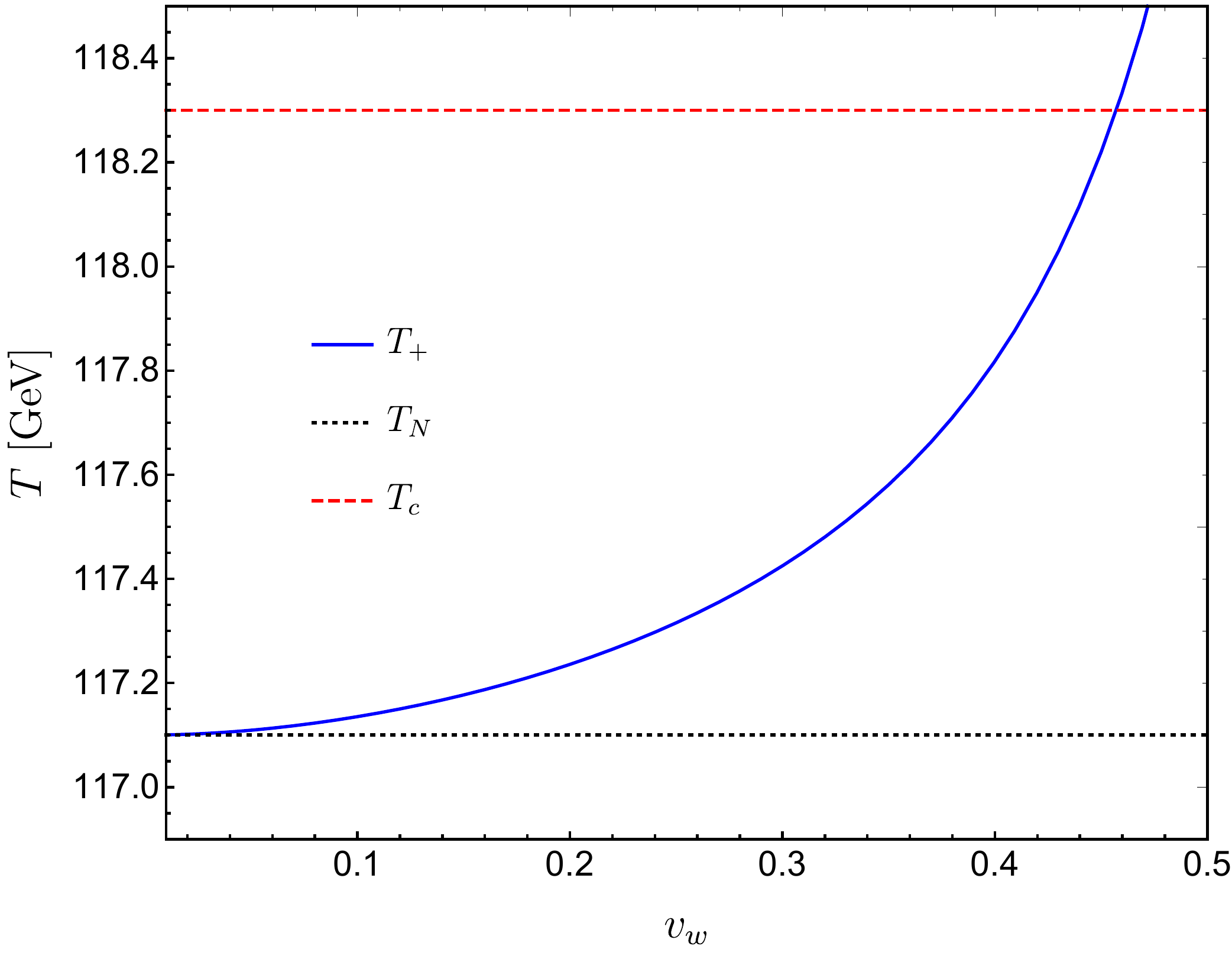}
	\caption{An illustration of temperature variation in front of the wall. We have $T_c=118.3$~GeV and $T_N=117.1$~GeV for the benchmark model parameters in the IDM.}\label{Tvariation}
\end{figure}

In Fig.\ref{Tvariation} we show an example of the heating effect with $T_c=118.3$~GeV and $T_N=117.1$~GeV. It can be seen that there is one value of $v_w$ at which $T_+$ will be larger than $T_c$
(However, in reality we think that the $T_+$ could not be larger than $T_c$. It could only approach $T_c$ because when $T_+ = T_c$ the potential difference is almost zero and the wall could not
be accelerated anymore.). We can give an approximation of the upper limit of the wall velocity \cite{Konstandin:2010dm},
\begin{equation}
	v_w \simeq \left(\frac{{\rm log} \frac{T_c}{T_N}}{6\alpha_c}\right)^{1/2}.
\end{equation} 
Actually, the heating effect can effectively be viewed as hydrodynamic backreaction because it effectively reduces the potential difference in front of and behind the wall~\cite{Konstandin:2010dm,Wang:2022txy}.

\section{Basic method of calculating bubble wall velocity}

Besides heavy particles in the standard model (top quark, $W$ and $Z$ bosons), in IDM we have extra heavy particle species, the $CP$-odd scalar $A$ and two charged scalars $H^{\pm}$ of the inert doublet $\eta$. As we can see in Eqs.~\eqref{mA} and \eqref{mH}, these three scalars have the same coupling such that same field-dependent mass and same interactions. Therefore, in this work one can treat them as the same species $A$. For the same reason, since $W^{\pm}$ and $Z$ have nearly degenerate mass, we can also treat them as the same species $W$. Other massive species that obtain light field-dependent mass should be almost in thermal equilibrium and we can treat them as background. 

In order to evaluate the friction term, we adopt the flow ansatz. We must know the out-of-equilibrium distribution part $\delta f$ for each massive particle population.
In principle, the  distribution function $f$ for the microscopic particles are described by the quantum Liouville equation. However, when the de-Broglie
thermal wavelength of particles in the system is smaller than the bubble wall thickness, the Wentzel-Kramers-Brillouin (WKB) condition $p \gg 1/L$ ($p$ is the momentum of concerned particle and $L$ is the bubble wall thickness) is satisfied. Then the background field varies slowly and hence the distribution function for each particle can be approximated by the following Boltzmann equation
\begin{equation}\label{bm}
	\frac{d}{dt}f = \left(\frac{\partial}{\partial t} + \dot{z}\frac{\partial}{\partial z} + \dot{p_z}\frac{\partial}{\partial p_z}\right)f = -C[f]\,\,,
\end{equation}
where $\dot{z}=p_z/E$ and $ \dot{p_z}=-\partial_z E=-(m^2)^{\prime}/(2E)$.
The distribution function for one species $f$, which deviates from its equilibrium form, can be expressed as
\begin{equation}
	f = \frac{1}{e^{(E+\delta)/T} \pm 1}\,\,,
\end{equation}
where $+$ is for fermions,  $-$ is for bosons, and 
$\delta$ is the perturbations ($\delta \ll 1$) that can be parametrized as \cite{Moore:1995si,Moore:1995ua}
\begin{equation}
	\delta = -\mu - \mu_{bg}  - \frac{E}{T}(\delta T + \delta T_{bg}) - p_z(\delta v +\delta v_{bg})\,\,.
\end{equation}
And we take the background chemical potential perturbation as zero \cite{Moore:1995si,Moore:1995ua}. We treat particles and antiparticles as one species neglecting $\mathrm{CP}$ violation.\footnote{Noting that this description is not appropriate for Infrared (IR) excitation with momenta $p \ll T$. These contributions can be important for boson species~\cite{Moore:2000mx,Kozaczuk:2015owa}.} $C[f]$ is the collision term, we will discuss the detailed calculations in next section.
The collision term and the EOM of the background field are model dependent. Then the Boltzmann equation can be written as 
\begin{eqnarray}
\begin{aligned}
	&\left(-f_{0}^{\prime}\right)\left(\frac{p_z}{E}\left[\partial_z \mu+\frac{E}{T} \partial_z\left(\delta T+\delta T_{b g}\right)+p_z \partial_z\left(\delta v+\delta v_{b g}\right)\right]+\partial_t \mu\right. \\
	&\left.+\frac{E}{T} \partial_t\left(\delta T+\delta T_{b g}\right)+p_z \partial_t\left(\delta v+\delta v_{b g}\right)\right)+T C[ \mu, \delta T, \delta v]=\left(-f_0^{\prime}\right) \frac{\partial_t\left(m^2\right)}{2 E}\,\,,
\end{aligned}
\end{eqnarray}
where $f_0$ is the equilibrium distribution function and $f_0^{\prime}=-\exp (E / T)/[\exp (E / T) \pm 1]^2$.
It is difficult to solve the full Boltzmann equation. 
However, one can truncate the full Boltzmann equation with three moments for an approximate solution. The three moments are chosen as $\int d^3 p /(2 \pi)^3, \int E d^3 p /(2 \pi)^2$, and $\int p_z d^3 p /(2 \pi)^3$~\cite{Moore:1995si,Moore:1995ua}.
Then, after the integration and keeping in mind that $\partial_tQ\rightarrow v_w Q'$, $\partial_zQ\rightarrow Q'$ for every heavy species $i = t, W, A$, we obtain
\begin{align}\label{per}
	v_w c_2^i\left(\mu_i^{\prime}+\mu_{b g}^{\prime}\right) &+v_w c_3^i\left(\delta T_i^{\prime}+\delta T_{b g}^{\prime}\right)+\frac{c_3^i T}{3}\left(\delta v_i^{\prime}+\delta v_{b g}^{\prime}\right)+\mu_i \Gamma_{\mu 1,i}+\delta T_i \Gamma_{T 1,i}=\frac{v_w c_1^i}{2 T}\left(m_i^2\right)^{\prime} \,\,,\nonumber\\
	v_w c_3^i\left(\mu_i^{\prime}+\mu_{b g}^{\prime}\right) &+v_w c_4^i\left(\delta T_i^{\prime}+\delta T_{b g}^{\prime}\right)+\frac{c_4^i T}{3}\left(\delta v_i^{\prime}+\delta v_{b g}^{\prime}\right)+\mu_i \Gamma_{\mu 2,i}+\delta T_i \Gamma_{T 2,i}=\frac{v_w c_2^i}{2 T}\left(m_i^2\right)^{\prime} \,\,,\nonumber\\
	\frac{c_3^i}{3} &\left(\mu_i^{\prime}+\mu_{b g}^{\prime}\right)+\frac{c_4^i}{3}\left(\delta T_i^{\prime}+\delta T_{b g}^{\prime}\right)+\frac{v_w c_4^i T}{3}\left(\delta v_i^{\prime}+\delta v_{b g}^{\prime}\right)+ \delta v_i T \Gamma_{v,i}=0\,\,,
\end{align}
with different integration coefficients of the collision term
\begin{align}\label{paracoll}
	\begin{split}
		&\int \frac{d^3 p}{(2\pi)^3 T^2} C[f_i]=\mu_i  \Gamma_{\mu1,i}+\delta T_i~ \Gamma_{T1,i}\,\,,\\
		&\int \frac{d^3 p}{(2\pi)^3 T^3} E C[f_i]=\mu_i \Gamma_{\mu2,i}+\delta T_i~ \Gamma_{T2,i}\,\,,\\
		&\int \frac{d^3 p}{(2\pi)^3 T^3} p_z C[f_i]=T\delta v_i   ~\Gamma_{v,i}\,\,.
	\end{split}
\end{align}
In Eqs.~\eqref{per}, the constants $c_j^{b/f}$ are defined as
\begin{equation}
	c_j^{b/f}T^{j+1} = \int E^{j-2}(-f_0')\frac{d^3p}{(2\pi)^3} \,\,.
\end{equation}
At lowest order in $m/T$, we have
\begin{equation}\label{cb}
	c_{1}^b = \frac{\log(2T/m_b)}{2\pi^2}\,\, ,\quad c_2^b = \frac{1}{6}\,\,,\quad c_3^b = \frac{3\zeta(3)}{\pi^2}\,\,,\quad c_4^b = \frac{2\pi^2}{15}\,\,,
\end{equation}
for bosons, whereas for fermions we have
\begin{equation}
	c_1^f = \frac{\log(2)}{2\pi^2}\,\,,\quad c_2^f = \frac{1}{12}\,\,,\quad c_3^f = \frac{9\zeta(3)}{4\pi^2}\,\,,\quad c_4^f = \frac{7\pi^2}{60}\,\,.
\end{equation}
$\zeta(x)$ is the Riemann zeta function.

Eqs.~\eqref{per} are not enough to solve for six quantities $\mu$, $\delta T$, $\delta v$, $\mu_{bg}$, $\delta T_{bg}$, $\delta v_{bg}$. 
Therefore, we need three more equations which describe the evolution of the background fluid \cite{Moore:1995ua}. The background fluid is the sum of all light particles which have negligible mass variation across the bubble wall.
All light particle species are treated as being at the same temperature $T + \delta T_{bg}$ and same velocity $\delta v_{bg}$ and the annihilation rates are fast enough that the background is in chemical equilibrium $\mu_{bg}=0$. Then we obtain three background equations which have a similar form as the heavy particles in Eqs.~\eqref{per}:
\begin{equation}
	\tilde{c}_4\left(v_w\delta T_{bg}' + \frac{\delta v_{bg}'}{3}T\right) = N_t(\mu_t\Gamma_{\mu2,t} + \delta T_t\Gamma_{T2,t}) + \sum_{\rm bosons}N_b(\mu_b\Gamma_{\mu2,b} + \delta T_b\Gamma_{T2,b})\,\,,\notag
\end{equation}
\begin{equation}\label{bg}
	\frac{\tilde{c}_4}{3}(\delta T_{bg}' + v_wT\delta v_{bg}') = N_tT \delta v_t\Gamma_{v,t} + \sum_{\rm bosons}N_bT \delta v_b\Gamma_{v,b}\,\,,\quad\mu_{bg} = 0\,\,,
\end{equation}
where we have used the fact that the collision term between massive particle species and light particle species in the fluid equations with opposite sign.
Here the heat capacity of the light degrees of freedom $\tilde{c}_4 = 78c_{4}^f + 19c_{4}^b$.

From Eqs.~\eqref{bg} the derivative of the background temperature and velocity can be written as:
\begin{align}
	\begin{split}\label{background}
	&\delta T_{bg}' = \frac{-v_w(\xoverline A + \xoverline B) + (\xoverline C+\xoverline D)}{\tilde{c}_4\left(1/3 - v_w^2\right)} \,\,,\\
    &\delta v_{bg}' = \frac{-3v_w(\xoverline C + \xoverline D) + (\xoverline A + \xoverline B)}{T\tilde{c}_4\left(1/3 - v_w^2\right)}\,\,,
	\end{split}
\end{align}
where
\begin{align}
	\begin{split}
		&\xoverline A = N_t(\mu_t\Gamma_{\mu2,t} + \delta T_t\Gamma_{T2,t})\,\,,\\
		&\xoverline B = \sum_{\rm bosons}N_b(\mu_b\Gamma_{\mu2,b} + \delta T_b\Gamma_{T2,b})\,\,,\\
		&\xoverline C = N_tT \delta v_t\Gamma_{v,t}\,\,,\\
		&\xoverline D = \sum_{\rm bosons}N_bT \delta v_b\Gamma_{v,b}\,\,.
	\end{split}
\end{align}

After substituting Eqs.~\eqref{background} and $\mu_{bg}=0$ into Eqs.~\eqref{per}, one can simplify the Boltzmann equation into the following compact matrix form
\begin{equation}\label{perturb}
	\hat A\delta'+\Gamma\delta =\Sigma\,\,,
\end{equation}
in our case the vector of perturbations comprise top quark $t$, vector bosons $W$ and scalar $A$,
\begin{equation}
	\mathbf{\delta}=(\mu_t,\delta T_t,T\delta v_t,\mu_W,\delta T_W,T\delta v_W,\mu_A,\delta T_A,T\delta v_A)\,\,,
\end{equation}
the source term
\begin{equation}\label{source}
	\Sigma =      
	 \frac{v_w}{2T}\left(c_1^t(m_t^2)',c_2^t(m_t^2)',0,c_1^W(m_W^2)',c_2^W(m_W^2)',0,c_1^A(m_A^2)',c_2^A(m_A^2)',0\right)\,\,,
\end{equation}
and
\begin{equation}
	\Gamma = \Gamma_0 + \frac{1}{\tilde{c}_4}\mathbb{M}
\end{equation}
\begin{equation}
	\hat A = \left(\begin{array}{ccc}
		\hat A_{t}&0 &0\\
		0&\hat A_{W}&0\\
		0&0&\hat A_{A}
	\end{array}\right),\quad \text{where}\quad
	\hat A_i = \left(\begin{array}{ccc}
		v_wc_2^i&v_wc_3^i&\frac{1}{3}c_3^i\\
		v_wc_3^i&v_wc_4^i&\frac{1}{3}c_4^i\\
		\frac{1}{3}c_3^i&\frac{1}{3}c_4^i&\frac{1}{3}v_wc_4^i\\
	\end{array}\right),
\end{equation}
\begin{equation}\label{gammamatrix}
	\Gamma_0 = \left(\begin{array}{ccc}
		\Gamma_t&0&0\\
		0&\Gamma_W&0\\
		0&0&\Gamma_A
	\end{array}\right),\quad\text{where}\quad
	\Gamma_i = \left(\begin{array}{ccc}
		\Gamma_{\mu1,i}&\Gamma_{T1,i}&0\\
		\Gamma_{\mu2,i}&\Gamma_{T2,i}&0\\
		0&0&\Gamma_{v,i}
	\end{array}\right)\,\,.
\end{equation}
Here $\mathbb{M}$ is
\begin{equation}
	\mathbb{M} = \left(\begin{array}{ccc}
		M_{tt}&M_{tW}&M_{tA}\\
		M_{Wt}&M_{WW}&M_{WA}\\
		M_{At}&M_{AW}&M_{AA}
	\end{array}\right),\quad\text{where}\quad
	M_{ij} = N_j\left(\begin{array}{ccc}
		c_{3}^i\Gamma_{\mu2,j}&c_{3}^i\Gamma_{T2,j}&0\\
		c_{4}^i\Gamma_{\mu2,j}&c_{4}^i\Gamma_{T2,j}&0\\
		0&0&c_{4}^i\Gamma_{v,j}
	\end{array}\right)\,\,,
\end{equation}
where $N_j$ is the total degree of freedom for the massive species ($N_t=12$, $N_W=9$ and $N_A=3$). 

We should note that when $m_b\gg T$ the approximation of $c_1^b$ in Eqs.~\eqref{cb} break down and we have
\begin{equation}
	c_1^b=(m_b/T)^{1/2}{\rm exp}(-m_b/T)/(2 \pi)^{3/2} \,\,,
\end{equation}
so that the contribution of the heavy boson particles to the right-hand side term in Eq.~(\ref{source}) is suppressed by a Boltzmann factor. Their contribution to friction would then be suppressed. For an extremely strong phase transition, this Boltzmann factor should be taken
into account and tends to increase the predicted velocity.

\section {Collision terms}
The collision terms are very important for calculating bubble wall velocity. If they are large enough, we expect that massive particles are only slightly apart from equilibrium and that they can not produce enough friction to prevent the bubble expansion. Then the wall velocity may be very large. On the other hand, if the collision terms are too small, our method that introduce perturbations
may be problematic; thus, it is essential to calculate collision terms accurately.

In most of the previous studies, only two massive species are considered; namely, the top 
quark and the $W$ bosons. Here, in our work we include extra massive particle species, the new scalars $H^{\pm}$ and $A$. We treat them as same species when calculating wall velocity so we can only calculate their collision terms once.

The collision term of the Boltzmann equation for species $i$ can be expressed as 
\begin{eqnarray}
\begin{aligned}
	C[f_i]=\frac{1}{2 \xoverline N_i} \sum_{} \frac{1}{2 E_p} \int \frac{d^3 k d^3 p^{\prime} d^3 k^{\prime}}{(2 \pi)^9 2 E_k 2 E_{p^{\prime}} 2 E_{k^{\prime}}} & \left|\mathcal{M}_{i }\left(p, k ; p^{\prime}, k^{\prime}\right)\right|^2(2 \pi)^4 \\ &\times \delta^4 \left(p+k-p^{\prime}-k^{\prime}\right)
    \mathcal{P}_{}\left[f_i(p)\right]\,\,,
\end{aligned}
\end{eqnarray}
where the sum is over all 4-body processes. The matrix elements are summed over helicities and colors of all four external quasiparticles, then divided by the number of degrees of freedom $\xoverline N_i$ corresponding to species $i$, 
\begin{eqnarray}
\xoverline N_t=\xoverline N_{\bar t}=6,\xoverline N_{W^+}=\xoverline N_{W^-}=\xoverline N_{Z}=3,\xoverline N_A=\xoverline N_{H^+}=\xoverline N_{H^-}=1\,\,.
\end{eqnarray}
$p$ is the momentum of the concerned heavy particle species, 
$k$ represents the momentum of the other incoming particle.
$p^{\prime}$ and $k^{\prime}$ denote the momenta of the outgoing particles. Then the Mandelstam variables $s$, $t$, $u$ are defined as 
$s=(p+k)^2=(p^{\prime}+k^{\prime})^2$,
$t=(p-p^{\prime})^2=(k-k^{\prime})^2$,
and  $u=(p-k^{\prime})^2=(k-p^{\prime})^2$. The population factor for process $ij \rightarrow mn$ is
\begin{equation}\label{pfactor}
\mathcal{P}_{}\left[f_i(p)\right] \equiv f_i(p) f_j (k) \left(1 \pm f_m(p^{\prime})\right)\left(1 \pm f_n(k^{\prime})\right)-f_m(p^{\prime}) f_n(k^{\prime})\left(1 \pm f_i(p)\right)\left(1 \pm f_j(k)\right)
\end{equation}
with the upper (lower) signs corresponding to bosons (fermions) and $f_i$ the appropriate perturbed distribution function for particle $i$, which we assume to take the form
\begin{equation}
f_i=\left(e^{\left(E+\delta_i\right) / T} \pm 1\right)^{-1} .
\end{equation}

We now analyze the distribution function at first order.
We can write $f_i = 1/(\exp a_i \pm 1)$ with $a_i = (E_i-\delta_i)/T$ in the plasma frame such that $1\pm f_i=f_i\exp a_i$.
Then Eq.~\eqref{pfactor} can be expressed as
\begin{equation}
	\mathcal{P}[f_i]=(e^{a_m+a_n} - e^{a_i+a_j})f_if_jf_mf_n\,\,,
\end{equation}
and we have
\begin{align}
	\begin{split}
		\exp(a_i + a_j) &= \exp\left[(E_i+E_j)/T\right]\times \exp(-\delta_i/T - \delta_j/T)\\
		&\simeq\exp\left[(E_i+E_j)/T\right]\times(1-\delta_i/T-\delta_j/T)\,\,.
	\end{split}
\end{align}
Then to the first order in $\delta$, we have
\begin{equation}
	\mathcal{P}[f_i]\simeq\frac{\delta_i+\delta_j-\delta_m-\delta_n}{T}f_{0,i}f_{0,j}(1\pm f_{0,m})(1\pm f_{0,n})\,\,,
\end{equation}
where $f_{0,i} = (e^{E_i/T} \pm 1)^{-1}$ is the equilibrium distribution for species $i$ and we use the fact that  $f_i e^{E_i/T}=f_{0,i} e^{E_i/T}+\mathcal{O}\left(\delta_i\right)=1 \pm f_{0,i}+\mathcal{O}\left(\delta_i\right)$. After integration we get the collision terms with the forms in Eqs.~\eqref{paracoll}.

Then we need to consider the dominant scattering processes of the massive particle species.
We calculate the collision terms by using the \emph{leading-log} approximation~\cite{Moore:1995si}: a) Neglecting masses of all the external particles; b) Neglecting $s$-channel contributions because they are not logarithmic; c) The logarithmic IR divergences are  regularized by the the thermal mass of the mediator. We use propagators of the
forms $1/(t-m_{i,T}^2)$ or $1/(u-m_{i,T}^2)$, where $m_{i,T}$ is thermal mass of the mediator.
The thermal masses for quarks and gluons are $m_{q,T}^2=g_s^2 T^2/6$ and $m_{g,T}^2=2g_s^2 T^2$, respectively.
$g_s$ is the strong coupling constant.
The thermal mass of $W$ boson  is $m_{W,T}^2= 11g_w^2 T^2/6$. And for the new scalars, $m_{A,T}^2=\lambda_3 T^2/24$. 
To see where the ``leading-log" comes from, we show  an example of the $t$-channel annihilation process $t\bar t \rightarrow gg$ with matrix element $-(64/9)g_s^4st/(t-m_{t,T}^2)^2$. The integral about $p^{\prime}$ and $k^{\prime}$ gives (in the center-of-mass frame),
\begin{eqnarray}
	\int \frac{p^{\prime 2} d p^{\prime} d \Omega_{p^{\prime}}}{(2 \pi)^3 4 E_{p^{\prime}} E_{k^{\prime}}} 2 \pi \delta\left(2 E_p-2 E_{p^{\prime}}\right) \frac{(2 p \cdot k) 2 p p^{\prime}\left(1-\cos \theta^{\prime}\right)}{\left[2 p p^{\prime}\left(1-\cos \theta^{\prime}\right)+m_{t,T}^2\right]^2}
	\simeq \frac{1}{8 \pi} \log \frac{2 p \cdot k}{m_{t,T}^2}\,\,,
\end{eqnarray}
where we have used the leading-log approximation that $E_{k^{\prime}}=E_{p^{\prime}}=\left|\vec{p}^{\prime}\right|=|\vec{k}^{\prime}|=p^{\prime} \text { and } E_k=E_p=|\vec{p}|=|\vec{k}|=p$. Here, $ 2 p \cdot k=2pk(1-$ $\cos \theta)+O\left(m_{t,T}^2\right)$, with $\theta$ being the plasma-frame angle between $\vec{p}$ and $\vec{k}$.
Then the remaining integrals about $\theta$ in the plasma frame contain
\begin{eqnarray}
	\int d \cos \theta \frac{1}{2} \log \left(\frac{2pk(1-\cos \theta)}{m_{t,T}^2}\right)=-1+\log \frac{4pk}{m_{t,T}^2}\,\,.
\end{eqnarray}
The remaining integrals can be done numerically.

However, in order to get more accurate results, we do the numerical integration directly like $\int \frac{d^3 p}{(2\pi)^3 T^2} C[f_i]$ by using the phase space parametrization  as discussed in Ref.~\cite{John:2000zq}. The Monte Carlo package VEGAS~\cite{Hahn:2004fe} is used in our numerical integration calculations. Then we can use Eqs.~\eqref{paracoll} to extract $\Gamma_{\mu 1}$, $\Gamma_{T 1}$, $\Gamma_{\mu 2}$, $\Gamma_{T 2}$ and $\Gamma_{v}$
for each particle species. We use another approximation in that we neglect the collisions between different massive species since we expect their contributions are subdominant compared with other processes. This is implied in Eq.~\eqref{gammamatrix}. The results of the matrix elements are shown in Table~\ref{table:ms}.  We only include processes at order of $g_s^4$, $y_t^2 g_s^2$, $g_w^2 g_s^2$, $g_w^4$, $\lambda_3^4$ as the dominant contributions. The matrix elements are summed over the helicities and colors of all four external states. 

We perform numerical integration and get,
for the top quarks,
\begin{eqnarray}
\begin{aligned}
	&\Gamma_{\mu 1, t} \simeq\left(5.0 \times 10^{-4} g_s^4+ 5.8\times 10^{-4} g_s^2 y_t^2\right) T \,\,,\\
	&\Gamma_{T 1, t} \simeq \Gamma_{\mu 2, t}\simeq\left(1.1 \times 10^{-3} g_s^4+ 1.3\times 10^{-3} g_s^2 y_t^2\right) T \,\,,\\
	&\Gamma_{T 2, t} \simeq\left(1.1 \times 10^{-2} g_s^4+ 4.0\times 10^{-3} g_s^2 y_t^2\right) T \,\,,\\
	&\Gamma_{v, t} \simeq\left(2.0 \times 10^{-2} g_s^4+ 1.8\times 10^{-3} g_s^2 y_t^2\right) T\,\,,
\end{aligned}
\end{eqnarray}
for the W bosons,
\begin{eqnarray}
\begin{aligned}
	&\Gamma_{\mu 1, W} \simeq\left(2.3 \times 10^{-3} g_s^2 g_w^2+ 2.0\times 10^{-3} g_w^4\right) T \,\,,\\
	&\Gamma_{T 1, W} \simeq \Gamma_{\mu 2, W} \simeq\left(4.7 \times 10^{-3} g_s^2 g_w^2+ 4.1\times 10^{-3} g_w^4\right) T \,\,,\\
	&\Gamma_{T 2, W} \simeq\left(1.5 \times 10^{-2} g_s^2 g_w^2+ 1.5\times 10^{-2} g_w^4\right) T \,\,,\\
	&\Gamma_{v, W} \simeq\left( 5.7 \times 10^{-2} g_s^2 g_w^2+ 1.5\times 10^{-2} g_w^4\right) T\,\,,
\end{aligned}
\end{eqnarray}
and for the new bosons $A$,
\begin{eqnarray}
\begin{aligned}
	&\Gamma_{\mu 1, A} \simeq 1.0 \times 10^{-2} \lambda_3^4 T \,\,,\\
	&\Gamma_{T 1, A} \simeq \Gamma_{\mu 2,A} \simeq 4.9 \times 10^{-3} \lambda_3^4 T \,\,,\\
	&\Gamma_{T 2, A} \simeq 5.1
	\times 10^{-3}\lambda_3^4 T \,\,,\\
	&\Gamma_{v, A} \simeq  1.8\times 10^{-3} \lambda_3^4 T \,\,.
\end{aligned}
\end{eqnarray}

\begin{table}[t]
	\centering
	\begin{tabular}{lc} 
		Process & $|\mathcal{M}_i|^2$  \\
		\hline \hline $\mathcal{O}\left(g_s^4\right):$ &  \\
		$t \bar{t} \leftrightarrow g g:$ & $\frac{128}{3} g_s^4\left(\frac{u}{t-m_{t,T}^2}+\frac{t}{u-m_{t,T}^2}\right)$  \\
		$t g \leftrightarrow t g:$ & $-\frac{128}{3} g_s^4 \frac{su}{\left(u-m_{g,T}^2\right)^2}+96 g_s^4 \frac{s^2+u^2}{\left(t-m_{t,T}^2\right)^2}$  \\
		$t q(\bar{q}) \leftrightarrow t q(\bar{q}):$ & $160 g_s^4 \frac{u^2+s^2}{\left(t-m_{t,T}^2\right)^2}$  \\
		$\mathcal{O}\left(y_t^2 g_s^2\right):$ &  \\
		$t \bar{t} \leftrightarrow h g,G^0 g:$ & $8 y_t^2 g_s^2\left(\frac{u}{t-m_{t,T}^2}+\frac{t}{u-m_{t,T}^2}\right)$ \\
		$t \bar{b} \leftrightarrow h G^{+} :$ & $8 y_t^2 g_s^2\left(\frac{u}{t-m_{t,T}^2}+\frac{t}{u-m_{b,T}^2}\right)$ \\
		$t g \leftrightarrow t h,t G^0:$ & $-8 y_t^2 g_s^2 \frac{s}{t-m_{t,T}^2}$  \\
		$t g \leftrightarrow b G^{+}:$ & $-8 y_t^2 g_s^2 \frac{s}{t-m_{b,T}^2}$  \\
		$t G^{-} \leftrightarrow b g:$ & $-8 y_t^2 g_s^2 \frac{s}{t-m_{t,T}^2}$  \\
		$\mathcal{O}\left(g_w^2 g_s^2\right):$ &  \\
		$Wq \leftrightarrow qg:$ &$-72 g_s^2 g_w^2 \frac{s }{t-m_{q,T}^2}$ \\
		$Wg \leftrightarrow q\bar{q}:$ &$-72 g_s^2 g_w^2 \frac{s }{t-m_{q,T}^2}$ \\
		$\mathcal{O}\left(g_w^4 \right):$ & \\
		$WW \leftrightarrow f \bar{f}:$ &$-\frac{27}{2} g_w^4 \left(\frac{3s}{t-m_{q,T}^2}+\frac{s}{t-m_{l,T}^2}\right)$\\
		$Wf \leftrightarrow Wf:$ &$360 g_w^4 \frac{u^2}{\left(t-m_{W,T}^2\right)^2}-\frac{27}{2} g_w^4 \left(\frac{3s}{u-m_{q,T}^2}+\frac{s}{u-m_{l,T}^2}\right) $  \\
		$\mathcal{O}\left(\lambda_3^4\right):$ &  \\
		$AA\leftrightarrow hh :$ &$ \frac{\lambda_3 ^4 v^4}{2}\left[\frac {1}{\left(t-m_{A,T}^2\right)^2} +\frac {1}{\left(u-m_{A,T}^2\right)^2} \right]$ \\
		$Ah\leftrightarrow hA :$ &$ \frac{\lambda_3 ^4 v^4}{2}\frac {1}{\left(t-m_{A,T}^2\right)^2} $ \\
		\hline
	\end{tabular}
	\caption{ Relevant 4-body processes and their corresponding matrix elements for massive particles in the leading-log approximation. The matrix elements are summed over the helicities and colors of all four external states. We only include processes at the order of $g_s^4$, $y_t^2 g_s^2$, $g_w^2 g_s^2$, $g_w^4$, $\lambda_3^4$ as the dominant contributions. Here $q$, $b$ and $l$ represent quarks, bottom quark and leptons respectively.}\label{table:ms}
\end{table}

\begin{figure}[htbp]
	\centering
	\subfigure{
		\begin{minipage}[t]{0.5\linewidth}
			\centering
			\includegraphics[scale=0.35]{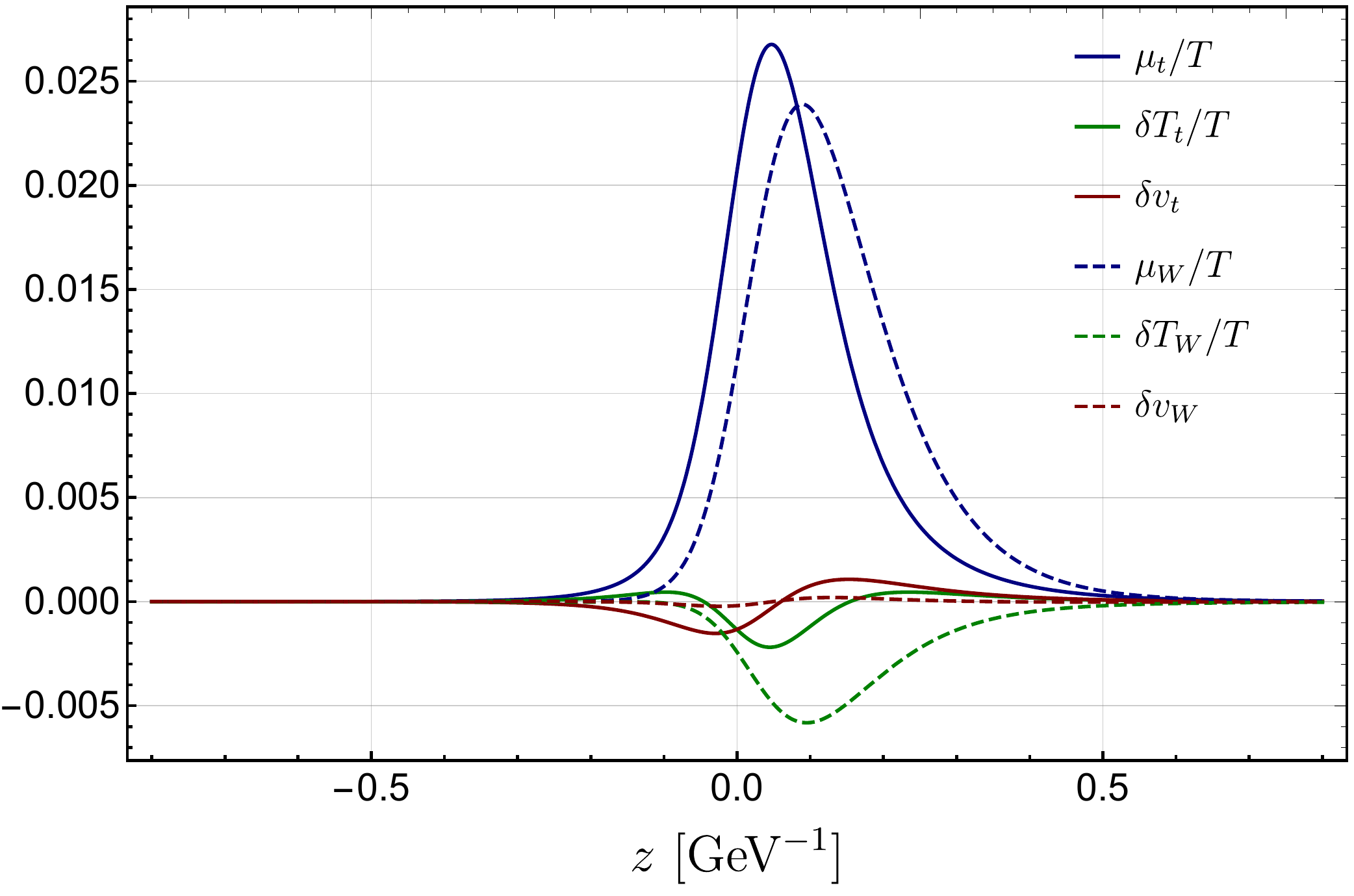}
	\end{minipage}}%
	\subfigure{
		\begin{minipage}[t]{0.5\linewidth}
			\centering
			\includegraphics[scale=0.35]{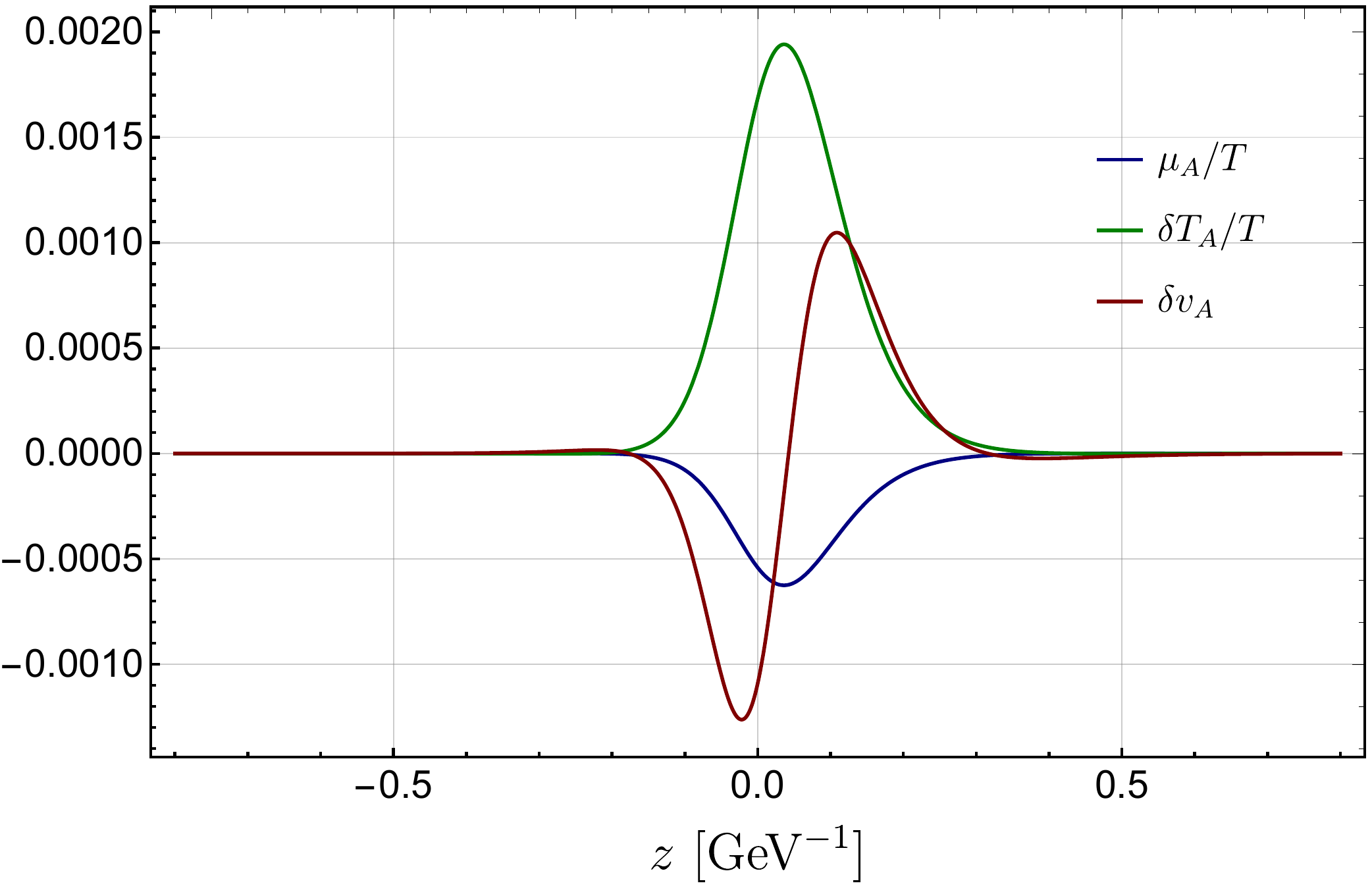}
	\end{minipage}}	
	\caption{Chemical, temperature and velocity perturbations of $W$, $t$ and $A$ for $v_w=0.1$ and $L = 0.1~\rm GeV^{-1}$. Left: perturbations for top quark (solid lines) and W boson (dashed lines). Right: perturbations for new scalars. }\label{perturbations}
\end{figure}

\section{Bubble wall velocity}

Given the above results, the perturbations could be derived from Eq.~\eqref{perturb} by using Green's function method,
\begin{equation}
	(\hat{A}^{-1} \Gamma)_{ij} \chi_{jk} =\chi_{jk}\rho_k\,\,,
\end{equation}
where $\rho_{k}$ are the eigenvalues of $\hat{A}^{-1} \Gamma$ and $\chi$ is the matrix constituting of the eigenvectors. Note that there is no sum on $k$. It is then straightforward to write down the Green's function
\begin{equation}
	G_i(z,y)=\mathrm{sgn}(\rho_i)e^{-\rho_{i}(z-y)} \Theta[\mathrm{sgn}(\rho_i)(z-y)].
\end{equation}
The Heaviside function means the boundary condition $\delta_{i}(z \rightarrow \pm \infty)=0$.
Then the perturbation $\delta_i$ will be given by
\begin{eqnarray}
	\delta_i(z)=\chi_{ij} \int_{-\infty}^{\infty} dy G_{j}(z,y)[\chi^{-1} \hat{A}^{-1} \Sigma(y)]_{j}\,\,,
\end{eqnarray}
where $\Sigma$ is given by Eq.~\eqref{source}. However, in order to evaluate this integration, we must know the explicit form of $m_i^2(\phi(z))$. To proceed further, we choose the ansatz of the bubble wall profile as
\begin{equation}
	\phi(z) = \frac{\phi_-}{2}\left(1 + \tanh\frac{z }{L}\right)\,\,,\label{bubble}
\end{equation}
where $\phi_-$ is the VEV of the Higgs boson in the broken phase at $T_-$, $\phi_-=\phi(T_-)$. $L$ is the bubble wall thickness.
This bubble wall profile might work well for relatively weak first-order phase transition.\footnote{For a SFOPT or the ultra supercooling phase transition, this profile ansatz might not be appropriate~\cite{Wang:2020jrd}.}
In this work, for the chosen benchmark parameters in the IDM, we just have a relatively weak first-order phase transition. Thus, it is reasonable to assume this shape in the following analysis.

\begin{figure}[htbp!]
	\centering
	\subfigure{
		\begin{minipage}[t]{0.5\linewidth}
			\centering
			\includegraphics[scale=0.355]{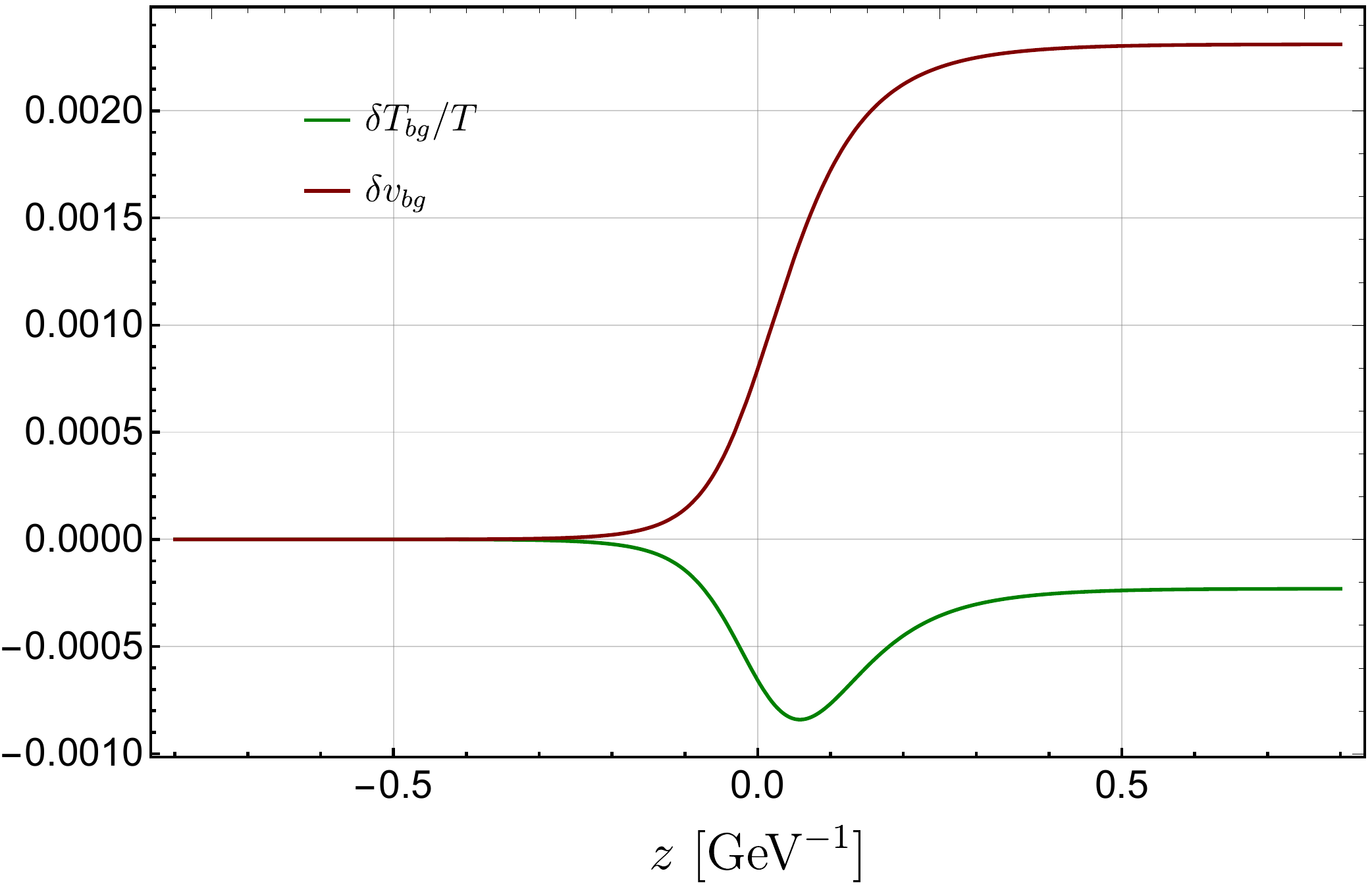}
	\end{minipage}}%
	\subfigure{
		\begin{minipage}[t]{0.5\linewidth}
			\centering
			\includegraphics[scale=0.39]{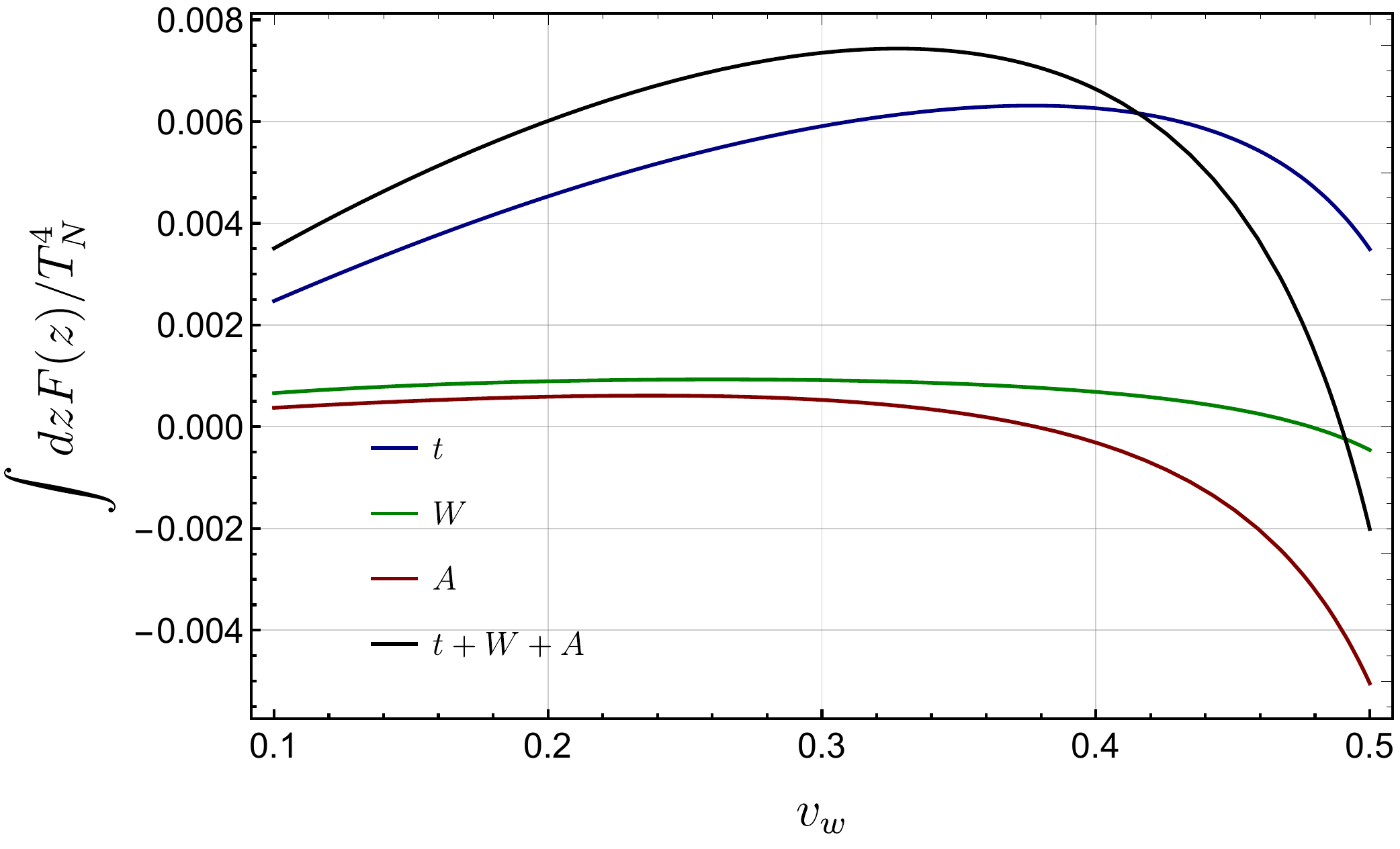}
	\end{minipage}}	
	\caption{Left panel: An illustration for the perturbation of background velocity and background temperature for $v_w=0.1,L=0.1~\rm GeV^{-1}$. Right panel: $\int dz F(z)/T_N^4$ for different species as a function of wall velocity $v_w$.}\label{ffff}
\end{figure}

Numerically solving the truncated Boltzmann equation, we could obtain the perturbations of all massive species as shown in Fig.~\ref{perturbations} and background perturbations in left panel of Fig.~\ref{ffff}, where we have chosen wall velocity $v_w=0.1$ and wall thickness $L = 0.1~\rm GeV^{-1}$. 
We can give brief discussions on the underlying physics of these numerical results. 
 One can see that the values of perturbations for $t$ and $W$ are larger than $A$'s. This is because the magnitudes of the collision rates of $A$ are larger than the rates of $t$ and $W$, such that $A$ is closer to equilibrium than $t$ and $W$. Also, for $t$ and $W$, their collision rates with respect to the chemical potential are smaller than those with respect to temperature, 
 so that $t$ and $W$ are easier to deviate from chemical equilibrium than thermal equilibrium. Then the extent of their decoupling from chemical equilibrium should be larger than thermal equilibrium. 
 Such that in Fig.~\ref{perturbations}, we can see that the perturbations of the chemical potential are much larger than temperature and velocity. It is the same for $A$, whose particle number-changing collision rate is larger than temperature-exchanging rate so the magnitude of chemical perturbations of $A$ is smaller than temperature and velocity. 
 As the wall is thick enough respect to the de-Broglie wavelength of massive particles, we expect that the shape of the dominant perturbations $\mu/T$ for $t$ and $W$ and $\delta T/T$ for $A$ will be closely proportional to $mm'$ and $\delta T_{bg}/T$ be proportional to $m^2$, which is consistent with our results.

Substituting the results obtained from Boltzmann equation under semiclassical and fluid approximations into Eq.~\eqref{eom3} we get that at leading-order perturbations, the EOM of Higgs can be approximated as
\begin{equation}\label{EOMp}
	\begin{aligned}
		S_{\mathrm{EOM}} \equiv	(1 & \left.-v_w^2\right) \phi^{\prime \prime}+\frac{\partial V_{\mathrm{eff}}\left(\phi, T_{+}\right)}{\partial \phi}+\frac{N_t T_{+}}{2} \frac{d m_t^2}{d \phi}  \times\left(c_1^t \mu_t+c_2^t\left(\delta T_t+\delta T_{b g}\right)\right) \\
		& +\sum_b \frac{N_b T_{+}}{2} \frac{d m_b^2}{d \phi}\left(c_1^b \mu_b+c_2^b\left(\delta T_b+\delta T_{b g}\right)\right)=0 \,\,,
	\end{aligned}
\end{equation}
where $N_i$ is the degree of freedom of particle $i$. Here $T_+$ is temperature just in front of the bubble wall and can be solved with the hydrodynamic treatment of expanding bubble which we have discussed previously.

Given specific bubble wall profile in Eq.~\eqref{bubble}, it is still difficult to fully solve the EOM in Eq.~\eqref{EOMp}. In practice, we could obtain an  approximate solution when the following two constraints are satisfied~\cite{Konstandin:2014zta}
\begin{equation}
	M_1=\int S_{\mathrm{EOM}}\phi'dz=0,\quad M_2=\int S_{\mathrm{EOM}}(2\phi-\phi_-)\phi'dz=0\,\,.\label{moments}
\end{equation}
The first equation $M_1=0$ in Eq.~\eqref{moments} means the total pressure on the bubble wall vanishes in the steady velocity regime.
And the second equation $M_2=0$ indicates the bubble wall thickness should not change anymore, which means the total pressure gradient should be zero for the steady bubble wall velocity.

The equilibrium, constant temperature part of Eq.~\eqref{moments} can be easily performed,
\begin{subequations}
	\begin{align}	
		\int \left[(1-v_{w}^2)\phi''+\frac{\partial V_{\rm eff}(\phi,T_{+})}{\partial \phi}\right]\phi' dz&=V_{\rm eff}(\phi_{-},T_{+})-V_{\rm eff}(\phi_+,T_{+}) \,\,,\\
		\int \left[(1-v_{w}^2)\phi''+\frac{\partial V_{\rm eff}(\phi,T_{+})}{\partial \phi}\right](2\phi-\phi_-)\phi' dz&=\frac{2(1-v_w^2)\phi_-^3}{15L^2}+\int_{\phi_+}^{\phi_-}\frac{\partial V_{\rm eff}(\phi,T_+)}{\partial \phi}(2\phi-\phi_-) d\phi\,\,,
	\end{align}	
\end{subequations}
The kinetic term $\phi''$ tends to stretch the wall (increasing $L$) while $V_{\rm eff}$ tends to accelerate and compress the wall.

Before further calculations, one should emphasize that the vacuum value $\phi_-$ using here is its value at $z\rightarrow \infty$. 
More precisely,  it is actually $\phi_-=\phi(T_-)$ that evaluated at $T_-$ instead of $\phi(T_+)$.\footnote{Generally $\phi(T_+)\neq \phi_+$ where $\phi(T_+)$ is the VEV at $T_+$ but $\phi_+$ is the field value in front of the bubble wall.}
This temperature jump is actually from the hydrodynamic effects discussed in Sec.\ref{ht}. However, as we can see from Eq.~\eqref{EOMp}, $\phi''$, the perturbations $\mu_i$, $\delta T_i$ and $v_i$ should vanish at $z\rightarrow \infty$. 
On the contrary, $\delta T_{bg}$ is not zero there. 
We will find that $\partial V/\partial \phi$ and the friction term coming from $\delta T_{bg}$ do not  exactly cancel with each other. 
 This behavior appears for the reason that the temperature jump is nonlinear so $T_- \neq T_+ +\delta T_{bg}$ and the vacuum value $\phi_0$ is very sensitive to temperature.
 
 We can see this more clearly in the following. Since we already had
\begin{equation}
	\frac{\partial V_{ \rm T}(\phi,T)}{\partial \phi} =\sum_{i}\frac{dm_i^2}{d\phi}\int\frac{d^3p}{(2\pi)^32E_i} f_{0,i}(p,T)\,\,,
\end{equation}
we simply expect that
\begin{equation}\label{VTbg}
	0 \equiv \frac{\partial V_{\rm eff}(\phi_-,T_-)}{\partial \phi_-} \simeq \frac{\partial V_{\rm eff}(\phi_-,T_+)}{\partial \phi_-}+ \frac{\partial^2 V_{\rm T}(\phi_-,T_+)}{\partial \phi_- \partial T_+}\delta T_{bg+}\,\,,
\end{equation}
where $\delta T_{bg+}=\delta T_{bg}(z\rightarrow \infty)$ is the background temperature perturbation at large positive $z$. The second term of Eq.~\eqref{VTbg} is just the friction term coming from $\delta T_{bg}$. However, in reality, the temperature varies nonlinearly in front and behind the wall, so this equality is too crude. In other words, Eq.~\eqref{EOMp} is not the real EOM at large positive $z$.

Actually, we can use $\partial V_{\rm T}/\partial T=-s$ such that
	\begin{equation}
		\frac{\partial^2 V_{\rm T}(\phi_,T_+)}{\partial \phi \partial T_+}\delta T_{bg}(z)=-\frac{\partial s(\phi,T_+)}{\partial \phi}\delta T_{bg}(z)\,\,,
	\end{equation}
which can be multiplied by $\phi'$ and integrated,
\begin{eqnarray}
	\int dz \phi'\left(\frac{\partial V_{\rm eff}(\phi,T_+)}{\partial \phi}- \frac{\partial s(\phi,T_+)}{\partial \phi} \delta T_{bg}(z)\right)&\simeq&V_{\rm eff}(\phi_-,T_+)-V_{\rm eff}(\phi_+,T_+)-s(\phi_-,T_+)\delta T_{bg+} \notag\\
	&+&\langle s \rangle \delta T_{bg+} \notag \\
	&\simeq& V_{\rm eff}(\phi_-,T_-)-V_{\rm eff}(\phi_+,T_+)+ \langle s \rangle \delta T_{bg+}\,\,,
\end{eqnarray} 
where the last term $\langle s \rangle \delta T_{bg+}$ can be approximated as $\langle s \rangle (T_--T_+)$ at first order which is consistent with Eq.~\eqref{Vs}.

So in order to minimize this inconsistency, we should find a new vacuum value $\phi_-$ to make EOM hold at infinity \cite{Friedlander:2020tnq}, 
\begin{equation}
	\left.\left(\frac{\partial V_{\mathrm{eff}}\left(\phi, T_{+}\right)}{\partial \phi}+\left(\frac{N_tT_+}{2}\frac{dm_t^2}{d\phi} c_2^t+\sum_b \frac{N_b T_+}{2}\frac{dm_b^2}{d\phi} c_2^b\right)\delta T_{bg}
	\right)\right|_{\phi=\phi_-, z \rightarrow \infty}=0 \,\,.
\end{equation}
The new $\phi_-$ does not minimize the potential but it cancels the friction term. Then we should recompute the perturbations and iterate.

Ref.~\cite{Friedlander:2020tnq} also introduced another method that one can find an $\mathcal{O}(1)$ parameter $y$ so that the modified EOM
\begin{eqnarray}
	\begin{aligned}
		(1-v_w^2)\phi''+ \frac{\partial V_{\rm eff}(\phi,T_+)}{\partial\phi} 
		&+ \frac{N_tT_+}{2}\frac{dm_t^2}{d\phi}\left(c_{1}^t\mu_t + c_{2}^t(\delta T_t + y \delta T_{bg})\right)\\
		&+ \sum_b \frac{N_b T_+}{2}\frac{dm_b^2}{d\phi}\left(c_{1}^b\mu_b + c_{2}^b(\delta T_b +y \delta T_{bg})\right) = 0
	\end{aligned}
\end{eqnarray}
is satisfied for larger positive values of $z$. In this work we will use the former method that redefines the VEV.

The last term of Eq.~\eqref{EOMp}, multiplied by $\phi '$, gives the friction term
\begin{equation}
	F(z)=\frac{N_tT_+}{2}\frac{dm_t^2}{dz}\left(c_{1}^t\mu_t + c_{2}^t(\delta T_t + \delta T_{bg})\right)+ \sum_b \frac{N_b T_+}{2}\frac{dm_b^2}{dz}\left(c_{1}^b\mu_b + c_{2}^b(\delta T_b + \delta T_{bg})\right)\,\,.
\end{equation}
Then the $M_1$ of Eq.~\eqref{moments} gives the condition of steady wall expansion,
\begin{equation}
	V_{\rm eff}(\phi_-,T_+)-V_{\rm eff}(\phi_+,T_+)=\int dz F(z)\,\,.
\end{equation}
We can see the behavior of $\int dz F(z)$ in Fig.~\ref{ffff}. We find that when $v_w$ approaches $c_s$ the friction force will approach  negative infinity which is due to the singularity of $\delta T_{bg}$, as seen in Eq.~\eqref{background}. This singularity may be caused by the selection of ansatz~\cite{Laurent:2020gpg,DeCurtis:2022hlx} or some physical reasons~\cite{Dorsch:2021ubz,Dorsch:2021nje,Laurent:2022jrs} but these works have shown that in the low-velocity regime our treatment still works well. Fortunately, we found that the velocity in the IDM is indeed small enough.

To do other integrals, it is much simpler to work in momentum space. However, it should be noted that $m_b=m_b(z)$ so the $c_1^b=c_1^b(z)$ is function of $z$. But we have noticed that the source term of perturbations is proportional to 
\begin{equation}
	\phi \phi'= \frac{\phi_-^2}{4}{\rm sech}^2 (\frac{z}{L})\left(1+{\rm tanh} (\frac{z}{L})\right)\,\,,
\end{equation}
which has a peak at $z=\frac{\rm{ln}2}{2}L$.  Therefore, the main contributions should come from $z$ around the peak. So it is reasonable to make an approximation that $c_1^b \simeq c_1^b(z=\frac{\rm{ln}2}{2}L)$.

Then the Boltzmann equations can be written in Fourier space as
\begin{eqnarray}
	ik\tilde{\delta_{i}}+(\hat A^{-1}\Gamma)_{ij}\tilde{\delta_{j}}&=&(\hat A^{-1}K)_{i}\widetilde{\phi \phi'} \,\,,\\
	ik\widetilde{\delta T}_{bg}&=&R_{i}\tilde{\delta_i}\label{Tbg} \,\,,
\end{eqnarray}
where $K$ and $R$ can be read from Eqs.~\eqref{source} and \eqref{background}, respectively.
\begin{gather}
	K=\frac{v_w}{2T}\left(c_1^t y_t^2,c_2^t y_t^2,0,\frac{c_1^W g_w^2}{ 2},\frac{c_2^W g_w^2}{2},0,c_1^A \lambda_3,c_2^A \lambda_3,0\right)\,\,,\\
	R=\frac{1}{\tilde{c}_4\left(1/3 - v_w^2\right)}\left(-v_w N_t \Gamma_{\mu2,t},-v_w N_t \Gamma_{T2,t},N_t \Gamma_{v2,t},-v_w N_W \Gamma_{\mu2,W},-v_w N_W \Gamma_{T2,W},\right. \notag\\
	\left. N_W \Gamma_{v2,W},-v_w N_A \Gamma_{\mu2,A},-v_w N_A \Gamma_{T2,A},N_A \Gamma_{v2,A}\right)\,\,.
\end{gather}

Substituting these into Eq.~\eqref{moments}, we get the contributions of perturbations to $M_1$ and $M_2$,
\begin{gather}
	\Omega_1=\int \frac{dk}{2 \pi}\left[f_i \chi_{ij}\frac{S_{j}}{\rho_{j}+ik}+\Upsilon R_{i}\chi_{ij}\left(\frac{S_{j}}{ik(\rho_{j}+ik)}-\frac{S_j \pi}{\rho_j}\delta(k)\right) \right]\widetilde{\phi \phi'}(k)\widetilde{\phi \phi'}(-k) \label{M1p}\,\,,\\
	\Omega_2=\int \frac{dk}{2 \pi}\left[f_i \chi_{ij}\frac{S_{j}}{\rho_{j}+ik}+\Upsilon R_{i}\chi_{ij}\left(\frac{S_{j}}{ik(\rho_{j}+ik)}-\frac{S_j \pi}{\rho_j}\delta(k)\right) \right]\widetilde{\phi \phi'}(k)\widetilde{2\phi^2 \phi'}(-k)-\phi_- \Omega_1 \,\,,\label{M2p}
\end{gather}
where $S=\chi^{-1}\hat A^{-1}K$, $f$ and $\Upsilon$ can be read from Eq.~\eqref{EOMp},
\begin{equation}
	f=T\left(\frac{c_1^t N_t y_{t}^2}{2},\frac{c_2^t N_t y_{t}^2}{2},0,\frac{c_1^W N_W g_w^2}{4},\frac{c_2^W N_W g_w^2}{4},0,\frac{c_1^A N_A \lambda_3}{2},\frac{c_2^A N_A \lambda_3}{2},0\right)\,\,,
\end{equation}
\begin{equation}
	\Upsilon=\frac{c_2^t N_t y_{t}^2}{2}+\frac{c_2^w N_W g_w^2}{4}+\frac{c_2^A N_A \lambda_3}{2}\,\,.
\end{equation}
It should be noted that we introduce Dirac delta function $\delta(k)$ in Eqs.~\eqref{M1p} and \eqref{M2p} because Eq.~\eqref{Tbg} can only determine $\delta T_{bg}$ up to a constant of integration. In order to satisfy the boundary condition $\delta T_{bg}(z\rightarrow -\infty)=0$, we must add this into the equation.

Using
\begin{gather}
	\widetilde{\phi \phi'}(k)=\frac{\phi_{-}^2}{2}(1-ikL/2)\frac{kL\pi}{2}\mathrm{csch}\frac{kL\pi}{2} \,\,,\\
	\widetilde{2\phi^2 \phi'}(k)=\frac{\phi_{-}^3}{12}(8-6ikL-k^2L^2)\frac{kL\pi}{2}\mathrm{csch}\frac{kL\pi}{2}\,\,,
\end{gather}
Eqs.~\eqref{M1p} and \eqref{M2p} can be solved by using the following integrals:
\begin{subequations}
	\begin{align}
		\int \frac{dk}{2\pi} \frac{1}{\rho + ik} \widetilde{\phi \phi'}(k)\widetilde{\phi \phi'}(-k)&=\frac{\phi_{-}^4}{16}\left[\left(\rho L-\frac{(\rho L)^3}{4}\right)I_{1}\left(\frac{\rho L \pi}{2}\right)+\frac{\rho L}{3}\right]\,\,,\\
		\int \frac{dk}{2\pi} \frac{1}{ik(\rho + ik)} \widetilde{\phi \phi'}(k)\widetilde{\phi \phi'}(-k)&=-\frac{1}{\rho}\frac{\phi_{-}^4}{16}\left[\left(\rho L-\frac{(\rho L)^3}{4}\right)I_{1}\left(\frac{\rho L \pi}{2}\right)+\frac{\rho L}{3}\right] \,\,,\\
		\int \frac{dk}{2\pi} \frac{1}{\rho + ik} \widetilde{\phi \phi'}(k)\widetilde{2\phi^2 \phi'}(-k)&=\frac{\phi_{-}^5}{12}\left[\left(\frac{\rho^4 L^4}{16}-\frac{\rho^3 L^3}{4}-\frac{\rho^2 L^2}{4}+\rho L\right)I_1(\frac{\rho L \pi}{2})\notag
		\right.\\ \left. -\left(\frac{\rho^2 L^2}{12}-\frac{\rho L}{3}-\frac{2}{5}\right)\right]\,\,,\\
		\int \frac{dk}{2\pi} \frac{1}{ik(\rho + ik)} \widetilde{\phi \phi'}(k)\widetilde{2\phi^2 \phi'}(-k)&=-\frac{1}{\rho}\frac{\phi_{-}^5}{12}\left[\left(\frac{\rho^4 L^4}{16}-\frac{\rho^3 L^3}{4}-\frac{\rho^2 L^2}{4}+\rho L\right)I_1(\frac{\rho L \pi}{2})\notag
		\right.\\ \left.-\left(\frac{\rho^2 L^2}{12}-\frac{\rho L}{3}\right)\right]\,\,,
	\end{align}
\end{subequations}
where
\begin{equation}
	I_{1}(a)=\int_{-\infty}^{\infty}dx\frac{x^{2}\mathrm{csch}^{2}x}{x^2+a^{2}}\,\,,
\end{equation}
which can be evaluated by contour integrals in the complex plane,
\begin{equation}
	I_{1}(a)=\frac{\pi |a|}{\mathrm{sin}^2 a}-2-\sum_{n=1}^{\infty}\frac{n(2\pi a)^2}{[(n\pi)^2-a^2]^2}\,\,.
\end{equation}

\begin{figure}[htbp]
	\centering
	\subfigure{
		\begin{minipage}[t]{0.5\linewidth}
			\centering
			\includegraphics[scale=0.5]{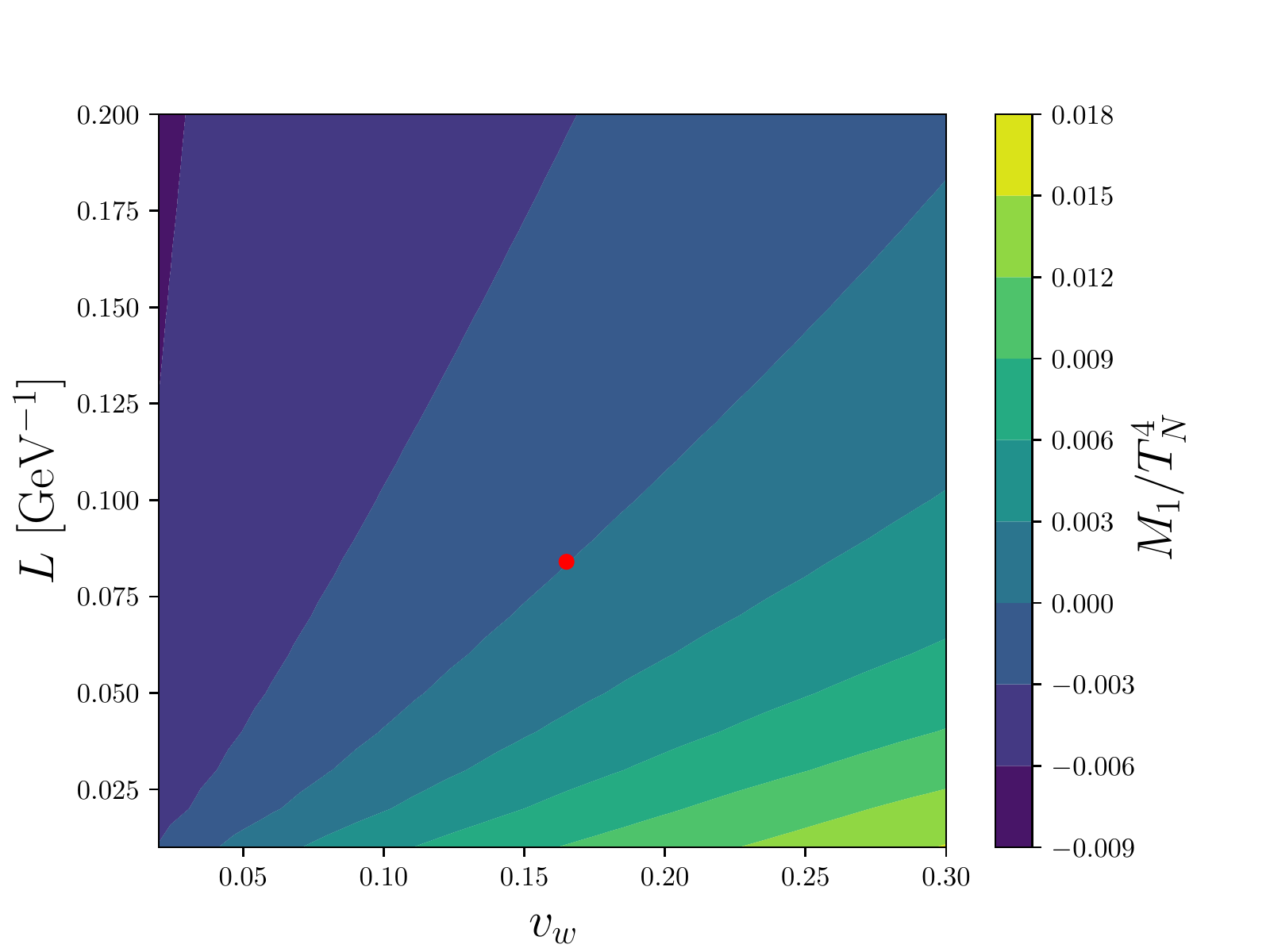}
	\end{minipage}}%
	\subfigure{
		\begin{minipage}[t]{0.5\linewidth}
			\centering
			\includegraphics[scale=0.5]{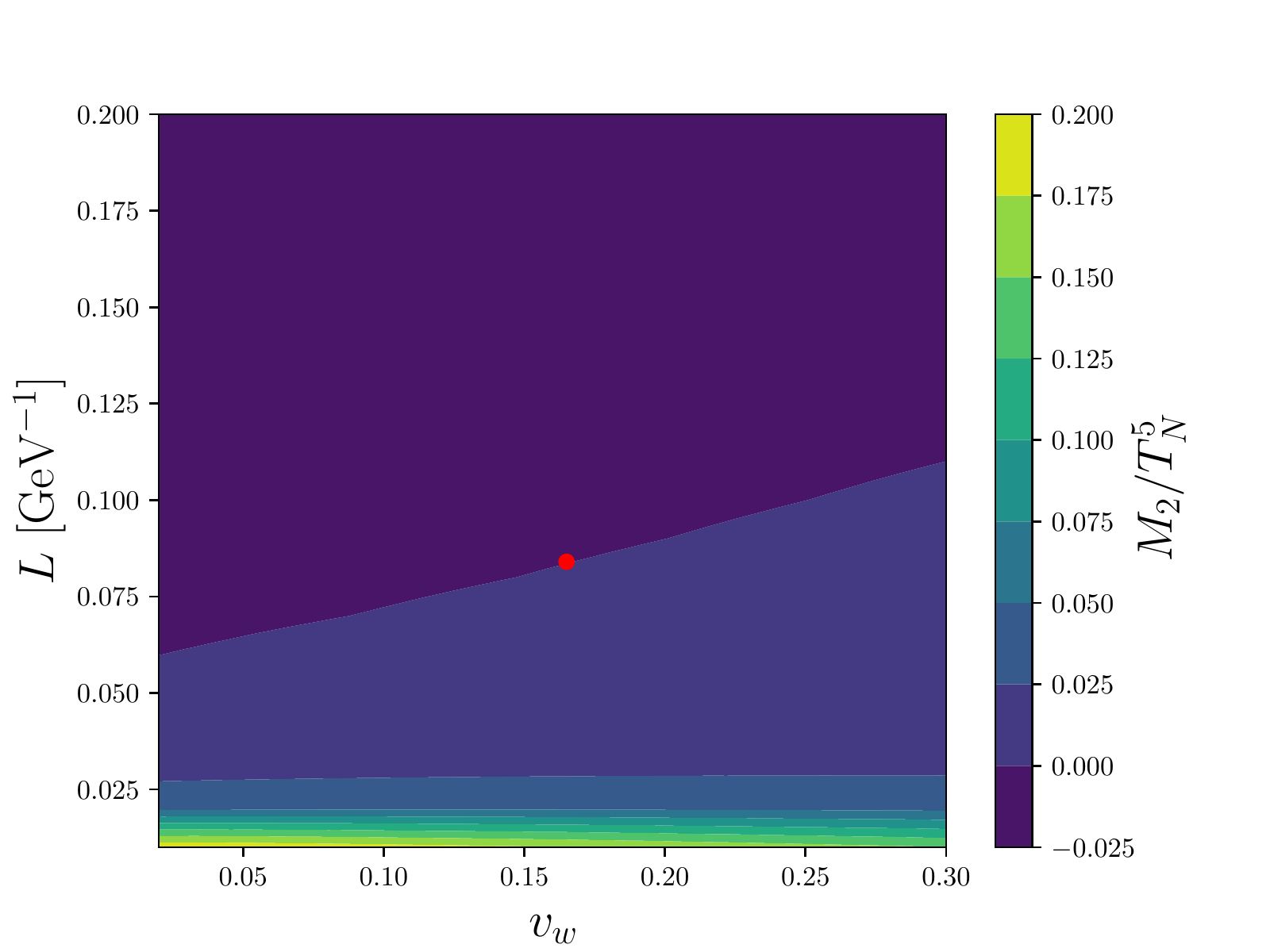}
	\end{minipage}}	
	\caption{Illustration of the two constraints $M_1/T_N^4$ and $M_2/T_N^4$. The red dots represent the final bubble wall velocity and bubble wall thickness for the benchmark parameters in IDM.}\label{constraint}
\end{figure}
In summary, the main procedure of calculating the bubble wall velocity is shown in Fig.~\ref{process}. For a given set of model parameters (in this case, the Benchmark A of IDM), we firstly calculate the thermodynamic quantities like the nucleation temperature $T_N$ and the phase transition strength $\alpha_N$ of the nucleation process with the ComoTransitions package. Then we perform a grid scan in the $(v_w, L)$ space. At each point in the grid we perform the next steps:

1. Solve the hydrodynamic equations of the SFOPT. This allows us to obtain the thermodynamic parameters evaluated in front of and behind the wall, that is $T_{+}, \alpha_{+}, v_{+}, \phi_+$, $T_{-}$, and $v_{-}$. Then we modify the field value by using the correct minimization conditions with the effective potential which should be evaluated at the temperature inside the bubble, with
\begin{equation}
	\left.\frac{\partial V_{\mathrm{eff}}\left(\phi,  T_{-}\right)}{\partial \phi}\right|_{\phi=\phi_-}=0 \,\,.
\end{equation}

3. Solve the truncated Boltzmann equations for concerned perturbations, which determines the friction force from the massive particles.

4. Choose new vacuum value $\phi_-$ such that it can satisfy the condition
\begin{equation}
	\left.\left(\frac{\partial V_{\mathrm{eff}}\left(\phi, T_{+}\right)}{\partial \phi}+\sum_i N_i \frac{d m_i^2}{d \phi} \int \frac{d^3 p}{(2 \pi)^3 2 E_i} \delta f_i(p, z)\right)\right|_{\phi=\phi_-, z \rightarrow \infty}=0\,\,.
\end{equation}

5. Recompute the perturbations and iterate. Then calculate $M_1$ and $M_2$.

In Fig.~\ref{constraint}, we show the two constraints $M_1$ and $M_2$ as functions of $v_w$ and $L$. For the two red dots,  $M_1=0$ and $M_2=0$ are satisfied simultaneously. The two dots correspond to the same value of bubble wall velocity and thickness. Therefore, the solution of two constraints is $v_w\simeq 0.165, L \simeq 0.084~\rm GeV^{-1}$ which implies that the wall velocity is subsonic in IDM . This solution is consistent with our assumption that $LT_N \gg 1$ such the WKB approximation holds in this case. 

In Table~\ref{vtable}, we also give the results of Benchmark B and C. We can see that in the allowed parameter spaces of IDM, the bubble wall velocity varies slightly around $v_w \simeq 0.165$. 

There are some uncertainties in our calculation. First, we parametrize the nonequilibrium with leading-order perturbations which may be problematic. Recently there were some modified schemes~\cite{Laurent:2020gpg,DeCurtis:2022hlx,Dorsch:2021ubz,Dorsch:2021nje,Laurent:2022jrs} but they have shown that in low-velocity regime the result of new scheme is almost the same as in the old scheme. Second,  we calculated collision terms by using leading-log approximation. Actually, if the collision term is larger, we expect that the velocity would be increased because the massive particles would be more closely to equilibrium. Then the friction would be decreased.

\begin{table}[t]
	\centering
	
	\setlength{\tabcolsep}{3mm}
	
\begin{tabular}{c|c|c|c|c}
		\hline\hline
		   &$T_c $~[GeV] &$T_N$~[GeV]&$v_w$& $L$~[$\rm GeV^{-1}$]\\
		\hline
		Benchmark A &118.3 &117.1 &0.165 &0.084 \\
		\hline
		Benchmark B &118.6 &117.5 & 0.164 & 0.085 \\
		\hline
		Benchmark C &119.4 & 118.4 & 0.164 & 0.088 \\
		\hline\hline
	\end{tabular}
\caption{Bubble wall velocity, bubble wall thickness and phase transition parameters for different benchmark points.}\label{vtable}
\end{table}
\section{Conclusion}

It is crucial and complicated to precisely calculate the bubble wall velocity, which is essential to the phase transition dynamics, the electroweak baryogenesis, the new dark matter mechanism from the bubbles, and the spectra of the phase transition gravitational wave. And more gravitational wave experiments (LISA, TianQin, Taiji,...) are proposed and they need accurate theoretic predictions on the gravitational wave signals. Thus, it becomes necessary and useful to calculate the model-dependent  bubble wall velocity both for the fundamental problems and the gravitational wave experiments~\cite{Azatov:2020ufh, Azatov:2021ifm}.

In this work, we have systematically calculated the bubble wall velocity in the well-motivated inert doublet model for the first time. 
The contribution from the heavy inert scalar bosons are taken into account. 
All the thermal masses are included in our numerical calculations. 
One difficulty is to consistently consider the hydrodynamic effects including the heating effects. To obtain the correct bubble wall velocity, it is important to figure out the correct temperature and vacuum value.
We consider the hydrodynamic effects to obtain more reliable results of the temperature and vacuum value.
Another difficulty is the precise calculations of the collision terms, which represent various particle scattering processes at the vicinity of the bubble wall so we have used the Monte Carlo algorithm to numerically obtain more precise collision terms. 
After having more rigorous hydrodynamic effects and collision terms, we get relatively more reliable nonequilibrium perturbations of various massive particles,
which are essential for the friction force.
Finally, by scanning the grid of wall velocity and wall thickness, we obtain the relatively precise 
bubble wall velocity in the inert doublet model for the first time. 
It is obviously smaller than the speed of sound and is favored by the traditional 
electroweak baryogenesis. This result could also help us to greatly reduce the uncertainty of the gravitational wave spectra from the strong first-order phase transition.

Our results might be useful for other similar models.  This procedure is also appropriate for other standard model-like models that have weak first-order phase transitions.\footnote{For strong enough phase transitions, it is necessary to introduce new methods to solve Boltzmann equations~\cite{Laurent:2020gpg,DeCurtis:2022hlx,Laurent:2022jrs}. 
}
Regarding the precise prediction of bubble wall velocity for a given new physics model, there are still some uncertainties from the collision terms and fluid approximations. More precise calculations on the collision terms~\cite{Arnold:2000dr,Arnold:2002zm,Arnold:2003zc,Bodeker:2017cim,Wang:2020zlf}
including the resummation over the large logarithmic terms~\cite{Hoche:2020ysm,Gouttenoire:2021kjv} and more accurate fluid ansatzes
are left for the future study.

\begin{acknowledgments}
We acknowledge valuable discussions with Eibun Senaha on the renormalization and daisy resummation schemes. The authors thank David Tucker-Smith, Marek Lewicki, Thomas Konstandin,  Glauber Carvalho Dorsch,  Benoit Laurent, Denis Werth and Carlos Tamarit for helpful correspondence.
This work is supported by the National Natural Science Foundation of China (NNSFC) under Grant No. 12205387.  X.W.  is supported by the China Postdoctoral Science Foundation under Grant No. 2022M713642.  
This work  is supported in part by  by Guangdong Major Project of Basic and Applied Basic
Research (Grant No. 2019B030302001).
\end{acknowledgments}

\bibliographystyle{apsrev}
\bibliography{velocity}

\end{document}